\documentclass[a4paper,fleqn,usenatbib]{mnras}

\usepackage{newtxtext,newtxmath}

\usepackage[T1]{fontenc}
\usepackage{ae,aecompl}

\usepackage{graphicx}	
\usepackage{amsmath}	
\usepackage{amssymb}	
\usepackage{bm}

\usepackage[dvipsnames]{xcolor}

\newcommand{\B}{\bm{B}}
\newcommand{\V}{\bm{v}}

\newcommand{\rp}{r_{\rm p}}

\newcommand{\Mp}{M_{\rm p}}
\newcommand{\wc}{w_{\rm c}}

\newcommand{\wwp}{w_{\rm p}}
\newcommand{\ffp}{f_{\rm p}}
\newcommand{\mpi}{{\rm \pi}}
\newcommand{\nup}{\nu_{\rm p}}

\title[Low mass planet dead zone migration - II]{Low mass planet migration in magnetically torqued dead zones -- II. Flow-locked and runaway migration, 
and a torque prescription}

\author[C.~P.~McNally et al.]{
Colin P.~McNally,$^{1,3}$\thanks{E-mail: c.mcnally@qmul.ac.uk (CPM)}
Richard P.~Nelson$^{1,3}$
and Sijme-Jan Paardekooper $^{1,2}$
\\
$^{1}$Astronomy Unit, School of Physics and Astronomy, Queen Mary University of London, London E1 4NS, UK\\
$^{2}$DAMTP, University of Cambridge, Wilberforce Road, Cambridge CB3 0WA, UK\\
$^3$Kavli Institute for Theoretical Physics, University of California Santa Barbara, CA 93106, USA\\
}

\date{Accepted XXX. Received YYY; in original form 2018 March 1}

\pubyear{2018}

\begin{document}
\label{firstpage}
\pagerange{\pageref{firstpage}--\pageref{lastpage}}
\maketitle

\begin{abstract}
We examine the migration of low mass planets in laminar protoplanetary discs, threaded by large scale magnetic fields in the dead zone that drive radial gas flows. As shown in Paper~I, a dynamical corotation torque arises due to the flow-induced asymmetric distortion of the corotation region and the evolving vortensity contrast between the librating horseshoe material and background disc flow. Using simulations of laminar torqued discs containing migrating planets, we demonstrate the existence of the four distinct migration regimes predicted in Paper~I. In two regimes, the migration is approximately locked to the inward or outward radial gas flow, and in the other regimes the planet undergoes outward runaway migration that eventually settles to fast steady migration.
In addition, we demonstrate torque and migration reversals induced by midplane magnetic stresses, with a bifurcation dependent on the disc surface density. We develop a model for fast migration, and show why the outward runaway saturates to a steady speed, and examine phenomenologically its termination due to changing local disc conditions.
We also develop an analytical model for the corotation torque at late times that includes viscosity, for application to discs that sustain modest turbulence. Finally, we use the simulation results to develop torque prescriptions for inclusion in population synthesis models of planet formation. 
\end{abstract}




\section{Introduction}

\defcitealias{2017MNRAS.472.1565M}{Paper I}

As a planet grows in a protoplanetary disc, angular momentum exchange between the planet and disc due to tidal torques begin to become important in driving migration when it exceeds the mass of Mars and approaches the Earth's mass. The theory of low mass planet migration has explicitly or implicitly been formulated in the context of viscous discs, by assuming that the underlying disc structure near the planet is smooth and unperturbed, and it has been supposed that the effective viscosity would be supplied by turbulence, which would also provide the accretion-driving stresses in the disc to account for mass flow onto the central star.

Our current understanding of the non-ideal magnetohydrodynamics of protoplanetary discs, however, suggests that 
they are likely not sufficiently turbulent to explain observed accretion rates, and are essentially laminar over large regions that incorporate areas of the disc traditionally associated with the sites of planet formation (i.e. between $\sim 0.1$--10~AU). The picture that emerges instead is one where disc accretion is driven by magnetised winds that are launched from thin ionised surface layers, and possibly also by large scale horizontal magnetic fields near the midplane, with the disc remaining laminar throughout the region described above
\citep{
2013ApJ...769...76B,
2013ApJ...772...96B,
2014ApJ...791...73B,
2014ApJ...791..137B,
2014A&A...566A..56L,
2015ApJ...801...84G,
2015ApJ...798...84B,
2015MNRAS.454.1117S,
2016ApJ...818..152B,
2016ApJ...821...80B,
2017A&A...600A..75B}.
In those discs where the vertical magnetic field is parallel to the angular momentum vector of the disc, the Hall effect, acting in the Ohmic dead zone \citep{1999MNRAS.303..239W,2008MNRAS.385.2269P}
can generate large scale radial and azimuthal fields, and a radial Maxwell stress, through the Hall-shear instability acting in concert with the Keplerian shear \citep{2008MNRAS.385.1494K,2013MNRAS.434.2295K}, 
even in a laminar flow \citep{2013ApJ...769...76B,2014A&A...566A..56L,2017A&A...600A..75B}.
In this paper, which follows on from \citetalias{2017MNRAS.472.1565M}, we aim to derive the properties of planetary migration torques in such discs, where we continue to focus on the Hall-modified region that drives an accretion flow throughout the vertical column of the disc and not just in its surface layers.

For planets which do not open a gap in the disc (i.e. the Type-I migration regime)\footnote{Although all planets eventually alter the surface density profile of the disc to some extent, the approximation that the surface density profile is unperturbed is a useful analytical approximation in understanding this limit.}, the migration torque consists of two basic components: the Lindblad torque, that arises due to the spiral wake driven by the planet, and the corotation torque, that originates from material close to the planet's orbit which undergoes horseshoe turns (or U-turns) in front of and behind the planet.
The Lindblad torque \citep{1979ApJ...233..857G} is insensitive to the presence or absence of viscous accretion stresses in the disc, and drives inward migration for most disc models. In the absence of other effects, the migration can be rapid, with migration times at 1~AU being $\sim {\rm few} \, \times 10^5$ years for an Earth mass planet and $\sim {\rm few} \times 10^4$ years for a 10 Earth mass body.

The corotation torque \citep[e.g.][]{1979ApJ...233..857G,1991LPI....22.1463W}, however, has much richer dependencies than the Lindblad torque, that in turn lead to a plethora of migration phenomena, some of which are the focus of this paper. The corotation torque can be viewed as resulting from the exchange of angular momentum with gas parcels undergoing horseshoe turns when encountering the planet. Clearly, there must be an asymmetry between the fluid elements making the turns in front of and behind the planet in order for a net torque to arise, and this is provided by a radial vortensity gradient in the disc, and perhaps also an entropy gradient if present \citep{2001ApJ...558..453M, 2008A&A...478..245P, 2008ApJ...672.1054B}. Furthermore, phase-mixing of material trapped on librating horseshoe orbits can remove the gradients that give rise to a net corotation torque, causing it to saturate and switch off. Viscous diffusion, however, can reestablish the vortensity gradient by promoting mixing between the corotation region and the surrounding disc, and hence can enable the corotation torque to be sustained. Under favourable disc conditions, with negative entropy and vortensity gradients and appropriate levels of viscous and radiative diffusion, a strong and positive corotation torque can be maintained that balances or exceeds the Lindblad torque, and hence prevents rapid migration of a planet into the central star \citep{2010MNRAS.401.1950P,2011MNRAS.410..293P,MassetCasoli2010}. We note that this corotation torque can be estimated at any point in time according to the instantaneous conditions in the disc at the location of the planet, and does not depend on the motion of the planet or on the history of the torques that have been applied to the planet or the disc. As such, we can describe it as a \emph{static} corotation torque.

A corotation torque can arise in an inviscid disc, but in this case the motion of the planet relative to the disc gas must be taken into account, resulting in a \emph{dynamical} corotation torque whose magnitude depends on the planet's motion \citep{2014MNRAS.444.2031P}. Here, the radial motion of the planet with respect to the disc causes distortion of the horseshoe region\footnote{In addition to the possibly small asymmetry arising from a vortensity gradient \citep{2009ApJ...703..845C}.}, which takes on a teardrop shape, with one wider and one narrower horseshoe turn, and the orbits narrowing in between. On the horseshoe turn where the outermost librating streamline is narrower, some background disc material will flow across the corotation region due to the radial motion of the planet, making only a single horseshoe turn before continuing on as part of the background disc. The mismatch between this material and the librating material, and the asymmetry of the librating streamlines, 
can lead to a corotation torque even in the absence of a turbulent viscosity to mix the vortensity of the librating region with the surrounding disc. For an inward migrating planet driven by Lindblad torques, this corotation torque can grow with time as the planet migrates and can eventually cause the inward migration to stall \citep{2014MNRAS.444.2031P}.

In \citet{2017MNRAS.472.1565M} (hereafter \citetalias{2017MNRAS.472.1565M}) we showed that in the non turbulent, very magnetically dead inner regions of a protoplanetary disc where radial flow of gas may be driven by stresses due to a Hall-generated laminar magnetic field, the resulting corotation torque can be described as a generalisation of the dynamical corotation torque. This generalisation involves accounting for the relative radial motion of the gas and planet.
In \citetalias{2017MNRAS.472.1565M}, we treated only torques acting on a planet on a fixed circular orbit with the gas flowing past the planet due to large scale magnetic accretion stresses. However, we were able to predict that there should be four distinct regimes of planet migration, that depend on the relative motion of the gas and the planet.  In this paper, we introduce live (fully dynamical) planets into our previous models and follow the longer term evolution of the net (Lindblad plus corotation) torque, allowing us to probe the four regimes of the dynamical corotation torque. Solutions where the planet is driven inwards, outwards, and where the net torque reverses are found. As with \citetalias{2017MNRAS.472.1565M}, we consider disc models with a simple barotropic isothermal equation of state, and hence we do not consider the influence of entropy gradients on the dynamical corotation torque, as have been considered previously by \citet{2015MNRAS.454.2003P} and \citet{2016MNRAS.462.4130P} in the context of viscous disc models. Hence, in this work we retain our focus on purely vortensity-related effects.

In summary, we expect that in all reasonable disc models, an embedded low mass planet will experience Lindblad torques that attempt to drive inwards migration. In a disc where the accretion stresses are primarily provided by turbulence, we expect that the corotation torques will be provided by the \emph{static} torques described above and in \citet{2011MNRAS.410..293P}. In a laminar magnetised disc, for which Hall EMFs are ineffective and accretion is driven only in narrow surface layers by the launching of a magnetised wind, such that there are no radial gas flows near the midplane, then we would expect the corotation torque to behave as the \emph{dynamical} torque discussed in \citet{2014MNRAS.444.2031P}. In a disc where the Hall effect plays an important role, such that accretion flows occur near the midplane and in the surface layers,  we would expect the corotation torque to behave according to dynamical torques described in this paper and in \citetalias{2017MNRAS.472.1565M}, with the long term evolution having a strong dependence on the relative motion between the planet and disc gas at the point when migration is initiated. 

The paper is organised as follows.
In Section~\ref{sec:basicanal} we discuss the basic analytical properties of the problem, define basic quantities and recapitulate relevant results from \citetalias{2017MNRAS.472.1565M}.
The first numerical experiments  involve enforcing specified initial migration to provoke the four basic regimes of migration described in \citetalias{2017MNRAS.472.1565M}  in Section~\ref{sec:specifiedmigration}.
We then explore reversing inward migration by magnetic disc torques in Section~\ref{sec:reversingmigration},
including a comparison between a magnetically driven inflow and a viscously driven one in Section~\ref{sec:viscous}.
In Section~\ref{sec:fastmigration} we analyse the fast migration regime both analytically and numerically, and present an interpolation table for the late-time torque in inviscid discs that we use in Appendix~\ref{sec:prescription} to construct a migration torque prescription for inclusion in N-body simulations of planet formation and population synthesis models. 
We propose and test an analytical model for steady-state torques which can exist when a residual viscosity is present in the disc in Section~\ref{sec:viscstedystates} and in Appendix~\ref{sec:viscmodel}.
We offer discussion of the results and their possible implications for planet formation scenarios in Section~\ref{sec:discussion},
and draw our conclusions in Section~\ref{sec:conclusions}.

\section{Fundamentals of Inviscid Type-I Migration}
\label{sec:basicanal}
This paper is concerned with the dynamical motion of a planet in an inviscid disc that is subject to a torque that drives a radial laminar flow through the full column density of the disc. As discussed in \citetalias{2017MNRAS.472.1565M}, such a torque can arise from horizontal magnetic fields that are generated by the Hall Shear Instability when the Hall effect is included in disc evolution calculations. We work in the limit where Type-I migration theory is valid, such that any change to the local surface density induced by the planet is small, and is not a primary factor in understanding the origin of the migration torque. Furthermore, we assume that the presence of the planet does not alter the background torque that acts on the disc. When considering a torque of magnetic origin, this is equivalent to assuming that the Ohmic resistivity is large enough to rapidly diffuse any magnetic field perturbations that are induced by the planet (as discussed below in Section~\ref{sec:characteristics}).

At the end of  \citetalias{2017MNRAS.472.1565M} we gave a formula for the corotation torque acting on a low mass planet in such a disc, assuming slow migration (the precise meaning of which will be discussed later) as
\begin{align}
\Gamma_{\rm hs} = 2\pi \left( 1-\frac{\wc (t)}{w(\rp)}\right) \Sigma_{\rm p} \rp ^2 x_s \Omega_{\rm p} \left[\frac{d \rp}{dt} -v_r \right]
\label{eq:unified}
\end{align}
where $w_{\rm c}(t)$ is the characteristic inverse vortensity of the librating streamlines trapped on horseshoe orbits,
$w(\rp)$ is the inverse vortensity of the background disc at the planet location,
$\Sigma_{\rm p}$ is the disc surface density at the planet position,
$\rp$ is the planet's radial position (such that $d \rp/dt$ is the migration speed),
$x_s$ is the half-width of the corotation region,
$\Omega_{\rm p}$ is the Keplerian orbital frequency at the planet position,
and $v_r$ is the radial velocity of the disc gas flow.
The vortensity, which is sometimes called the specific vorticity, can be written as $\omega= (\nabla \times \V)/\Sigma$.
As the flow is two dimensional, the curl of the two dimensional velocity 
only has a $z$-component so without ambiguity the vortensity can be treated as a scalar.
Even if the migration is fast, in that the planet moves radially quickly with respect to the gas, 
then we expect the torque to have the 
same sign as suggested by this form. Areas of parameter space close to  where the torque 
changes sign are in the slow migration regime.

In \citetalias{2017MNRAS.472.1565M} we identified four regimes of behaviour in this torque expression separated by sign changes of two quantities. These regimes correspond to \emph{qualitative differences} in the expected migration behaviour of a planet. We restrict our attention in this paper to Keplerian discs with radially decreasing vortensity profiles. Since vorticity $\omega=\Omega/(2\Sigma)$, where $\Omega$ is the Keplerian orbital angular velocity, our power-law disc models have $\Sigma \propto r^{-\alpha}$ where $\alpha<3/2$.
As equation~(\ref{eq:unified}) applies formally when the relative radial velocity of the gas and planet are small and the vortensity contrast is small, these parameters determining the borders between regimes are the disc gas radial flow velocity $v_r$ and the `initial' planet migration velocity $d \rp/dt$ close to transitions between regimes.
Given this, the four regimes and the expected long-term migration behaviours are:
\begin{description}
\item[ (i) $v_r \leq  0$ and {$[d \rp/dt - v_r] <0$}:]
  The disc accretion flow and planet migration are inwards, and the planet initially migrates faster than the disc flow $\rightarrow$ The planet migrates inwards, close to but slightly faster than the gas inflow speed.
\item[ (ii) $v_r < 0$ and {$[d \rp/dt - v_r]>0$}:] 
  The disc accretion flow and planet migration are inwards, and the planet initially migrates slower than the disc flow $\rightarrow$ The planet's inward migration reverses and runs away outwards.
\item[ (iii) $v_r > 0$ and {$[d \rp/dt - v_r]<0$}:] 
  The disc flow is outward and the planet initially migrates inwards $\rightarrow$ The planet's inward migration reverses but the outward migration speed cannot exceed that of the gas.
  \item[ (iv) $v_r > 0$ and {$[d \rp/dt - v_r]>0$}:]
  The disc flow and planet migration are outwards, and the planet is initially migrating faster than the disc flow $\rightarrow$  The planet's outward migration can run away.
\end{description}
Regimes (i) and (iii) result in slow migration, by which we mean that the planet asymptotically migrates with the radial motion of the gas. Hence, the planet's motion is locked to the radial gas flow.
Regimes (ii) and (iv), however, result in the planet moving fast with respect to the gas.
The meaning of fast and slow radial motion is defined in terms of the relevant flow timescale across the planet's corotation region.

To describe the speed of the radial flow of gas driven by the background torque induced by a laminar magnetic field, we use 
the ratio of the flow crossing time across the corotation region to the libration time
\begin{align}
\chi \equiv \frac{\tau_{\rm f}}{\tau_{\rm lib}},
\end{align}
where $\tau_{\rm f}$ is the flushing timescale, and $\tau_{\rm lib}$ is the libration timescale.
In turn, the flushing timescale is
\begin{align}
\tau_{\rm f} \equiv \frac{2 x_s}{-v_r},
\end{align}
where $x_s$ is the width of the corotation region, and $v_r$ is the radial velocity of the disc gas flow,
and the libration timescale is given as
\begin{align}
 \tau_{\rm lib}  = \frac{4\pi}{x_s |d\Omega/dr|}.
\end{align}
Hence, a positive value of $\chi$ corresponds to a negative background torque driving gas inflow.
Finally, in these expressions the half-width of the corotation region can be approximated as 
\begin{align}
x_s \approx 1.2 \rp \sqrt{q/h}\ ,
\label{eqn:x_s}
\end{align}
when the disc is two-dimensional and the planet potential is smoothed as a Plummer sphere with smoothing length $b=0.4H$,
with $q$ the planet-star mass ratio $q=\Mp/M_\star$,
$\rp$ being the radial position of the planet, $H$ the disc scale height 
and $h=H/r$, following \citet{2006ApJ...652..730M}.

The analytical model of \citet{2014MNRAS.444.2031P}  was constructed in the slow migration regime,
where the time taken for the planet to migrate across its own corotation regime is long compared to the libration period,
that is 
\begin{align}
\left| \frac{x_s}{d \rp /dt} \right| \gg \tau_{\rm lib}\, .
\end{align}
In \citetalias{2017MNRAS.472.1565M} the analogous condition for slow radial gas flow was $| \chi | \gg 1$ 
(we are slightly more careful here to explicitly write the absolute value bars when discussing this condition).
Thus, the unified form equation~(\ref{eq:unified}) is only strictly valid when
\begin{align}
\left| \frac{x_s}{d \rp /dt - v_r}\right| \gg \tau_{\rm lib}\, ,
\end{align}
which gives the earlier promised definition of slow migration.
However, freely migrating planets will often migrate at rates which violate this condition, particularly in regime (ii) or regime (iv).
We address this fast migration in Section~\ref{sec:fastmigration}.
It is also convenient to define an appropriately generalised version of the $\chi$ parameter 
taking into account the radial motion of the planet as
\begin{align}
\chi_{\rm G} &\equiv \frac{3 x_s^2 \Omega_p}{4 \pi \rp (d\rp/dt-v_r)}.
\end{align}
Again $\chi_{\rm G}>1$ corresponds to slow migration.

A second property that immediately becomes apparent from an examination of Equation~(\ref{eq:unified}) is that there exists a bifurcation in the behaviour of planets attempting to change migration from regimes (i) to (ii). This could occur, for example, if a planet is in regime (i) with relatively slow gas inflow, but moves into a region of the disc where the inflow is faster. Or if the gas inflow changes with time and speeds up such that it becomes faster than the planet's migration speed. The bifurcation occurs because changes to the factor $\left( 1-{\wc (t)}/{w(\rp)}\right)$ in Equation~(\ref{eq:unified}) that occur during the regime (i) phase of migration are reversed as the disc torque and inflow speed of the gas are increased and the planet tries to enters regime (ii). This unwinding of the relative inverse vortensity perturbation causes the corotation torque, $\Gamma_{\rm hs}$, to pass through zero, and at that moment the migration speed of the planet is only determined by the Lindblad (wake) torque. Hence, for a particular value of $v_r$, a critical surface density separates scenarios where planet migration can transition from regime (i) to regime (ii), since the migration speed at the point of change over depends linearly on the surface density.  To be able to make the transition from regime (i) to regime (ii) the Lindblad torque must drive the planet radially 
inwards slower than the disc gas inflow velocity $v_r$, or 
\begin{align}
\left. \frac{ d \rp}{dt}\right|_{\rm crit} = v_r\ .
\end{align}
The migration rate $d\rp /dt$ can be expressed in terms of the migration torque $\Gamma$ as
\begin{align}
 \frac{ d \rp}{dt} &= 2 \rp \frac{\Gamma/\Gamma_0}{\Mp \sqrt{G M_\star \rp}}\Gamma_0\, ,
\end{align}
where $\Mp$ is the planet mass, $G$ is the gravitational constant, $M_\star$ is the mass of the central star, and $\Gamma_0$ is a scale used to nondimensionalized the torque.
A linear calculation of the Lindblad torque in a more general adiabatic disc yields
\begin{align}
\gamma \Gamma_{\rm L}/\Gamma_0 = -(2.5 + 1.7 \beta - 0.1 \alpha) \left(\frac{0.4}{b/h}\right)^{0.71} 
\end{align}
for the rest of the dependence
where $\gamma$ is the adiabatic index of the gas ($\gamma=1$ for isothermal gas),
$\beta$ is the negative power law slope of the radial temperature gradient, and $\alpha$ is the negative power law slope of the radial surface density profile of the disc,
 $h=H/r$ is the aspect ratio of the disc, and $b$ is the smoothing length of the planet potential (for a Plummer sphere softening)
\citep{2008A&A...485..877P}.
The only dependence of the migration rate on surface density at the planet position $\Sigma_{\rm p}$ is contained in the parameter nondimensionalizing the Lindblad torque
\begin{align}
\Gamma_0 = \left(\frac{q}{h}\right)^2 \Sigma_{\rm p} \rp^4 \Omega_{\rm p}^2,
\end{align}
where  $\Omega_{\rm p}$ is the angular frequency of the planet's orbit.
Thus in a globally isothermal disc  with a laminar magnetic inflow torque acting on the gas 
there exists a $\Sigma_{\rm p,crit}$ above which a planet cannot cross from regime (i) to regime (ii).
This phenomena will be demonstrated in Section~\ref{sec:reversingmigration},
where the critical surface density sets a divide between planets which can reverse their migration and migrate outwards,
and those which are driven inwards in the same disc conditions.

\subsection{Characteristics of the problem established in Paper I}
\label{sec:characteristics}
Here we discuss other aspects of the problem established previously.
Importantly, in \citetalias[][sections 2.1.1 and 3.2]{2017MNRAS.472.1565M}, by means of ionisation chemistry calculations, we found that in the Ohmic resistivity
 dominated dead zone of a protoplanetary disc, the Ohmic diffusion timescale for the magnetic field is much faster 
then the U-turn flow timescale for low mass planets.
This implies that the deviations from Keplerian flow due to the presence of the planet are not expected to result in 
significant changes to the magnetic field configuration, and that hence simulations including the live evolution of the 
magnetic field will be equivalent to one including only a fixed field or simply the equivalent Lorentz force from the 
static magnetic field given in the initial condition.

As a consequence of this fact, though the models in this work are constructed with a magnetic field strength 
and a resistivity value, these are degenerate, and the only physical parameter is the radial flow velocity driven 
by the resulting magnetic field configuration.
Significant midplane flow velocities can be included within the total accretion rate typically observed onto
protoplanetary disc hosting stars \citepalias{2017MNRAS.472.1565M}. 
Thus our models are stated in terms of this radial flow velocity, and not the underlying
magnetic field strengths and resistivities. This allows our two dimensional models to be compared to three dimensional 
configurations with the same radial flow velocity.

As in \citetalias{2017MNRAS.472.1565M} the models in this work are two dimensional. 
The corotation region for a low mass planet has been previously found to 
 to have a columnar flow structure in three dimensions \citep{2016ApJ...817...19M}.
Thus, the results for two dimensional models of the corotation torque 
should be similar to three dimensional models.
In this work we again consider only an isothermal disc to isolate the effects of the vortensity related corotation torques. 
We further discuss the possible effects of realistic thermodynamics in Section~\ref{sec:discussion}.

\section{Demonstrating Four Migration Regimes}
\label{sec:specifiedmigration}
To demonstrate that the four migration regimes described above and predicted in \citetalias{2017MNRAS.472.1565M} exist, in this section
we present customised experiments with forced initial planet migration and disc parameters such that reasonably clean 
examples of all four regimes are produced.
In a disc with a fixed radial inflow,
 we force the initial inward migration rate, either a little slower than the disc inflow, or much faster.
This excites regime (i) or regime (ii) behaviour once the planet is released and allowed to migrate according to the disc torque.
A second set of experiments with disc outflow excites regime (iii) by forcing initial inward migration, which results in the planet turning around and migrating outwards, 
with the planet asymptotically migrating at close to the disc gas outflow speed.
Regime (iv) is excited by initially specifying rapid outward migration, and outward fast migration is maintained and increases in speed slightly when the planet is released.

We adopt the two dimensional disc model defined in \citetalias{2017MNRAS.472.1565M}.
This is a globally isothermal disk in cylindrical coordinates $(r,\phi)$ with surface density
$\Sigma = \Sigma_0 \left(r/r_0 \right)^{-\alpha}$ where $r_0$ is a reference radius and $\alpha$ is a constant determining 
the power-law slope of the disc surface density profile.
Our disks are always globally isothermal, with the vertical scale height aspect ratio $h=0.05$ at radius $r_0=1$,
which also specifies the constant isothermal sound speed $c_s$.
Throughout this paper, when times are given in units of orbits, they are orbits at $r_0=1$.
All models are presented with a central stellar mass of $M_\star = 1$ and gravitational constant $G=1$.
The planet mass  is denoted  $\Mp$ but usually specified in terms of the planet-star mass ratio $q=\Mp/M_\star$.
Our simulations are performed with FARGO3D \citep{2015ascl.soft09006B,2016ApJS..223...11B} in two dimensions on a cylindrical grid with linear spacing in radius,
using a body force to drive radial gas flow instead of a magnetic field. 
The domain extends in azimuth $2\pi$ and in radius from $0.3$ to $1.7$, and the fiducial resolution used throughout is $(N_r,N_\phi) = (1024,2048)$.
As in \citetalias{2017MNRAS.472.1565M}, modified \citet{2006MNRAS.370..529D} style damping zones at the boundary are used with the 
inner one having a radial width of $0.1$ and the outer a radial width of $0.2$.
The body force added to the momentum equation is the Lorentz force which would be produced by a spiral magnetic field given by
\begin{align}
\B &= B_0 \left(\frac{r}{r_0}\right)^{-1} \bm{\hat{r}}  -2 B_0 \Omega_0 r_0^2 \frac{\mu_0}{\eta} \left(\frac{r}{r_0}\right)^{-1/2} \bm{\hat{\phi}} 
\end{align}
where $B_0$ is a constant with units of the magnitude of magnetic field, $\Omega_0$ is the 
Keplerian angular velocity at $r_0$, $\mu_0$ is the magnetic permeability of free space, and $\eta$ is the Ohmic resistivity of the disc gas.
The velocity considered in calculating this force is the Keplerian velocity, as this is the thin disc 
approximation used in constructing the spiral magnetic field in  \citetalias{2017MNRAS.472.1565M},
and the presence of the planet will not significantly alter the magnetic field configuration as demonstrated in  \citetalias{2017MNRAS.472.1565M}.
This force produces a radial flow in the two-dimensional disc with velocity 
\begin{align}
v_r &= -\frac{2 B_0^2 r_0^2}{ \eta \rho L_z r}\, ,\label{eq:vr}
\end{align}
where $\Sigma =  \rho L_z$, $\rho$ is the midplane density, and we take $L_z$ to be a constant in these models.
As in \citetalias{2017MNRAS.472.1565M}, using just this force and eliminating the induction equation from the model
 is possible as the disc has sufficient Ohmic diffusivity that the shape of the 
magnetic field lines is not significantly altered by the flow perturbations due to the planet.
We initially specify the disc gas velocity with $v_r=0$, so the model is run for $10$ orbits to relax the radial flow to its 
equilibrium state before adding the planet potential.
Self-gravity of the disc gas is neglected, and the axisymmetric component of gravity is thus neglected when calculating the 
planetary migration torque to ensure consistency between the angular velocity of the planet and the disc gas \citep{2008ApJ...678..483B}.
The entire disc mass, including that in the Hill sphere (which is not bound to a low mass planet) is included in the torque calculation.

\begin{figure}
\includegraphics[width=\columnwidth]{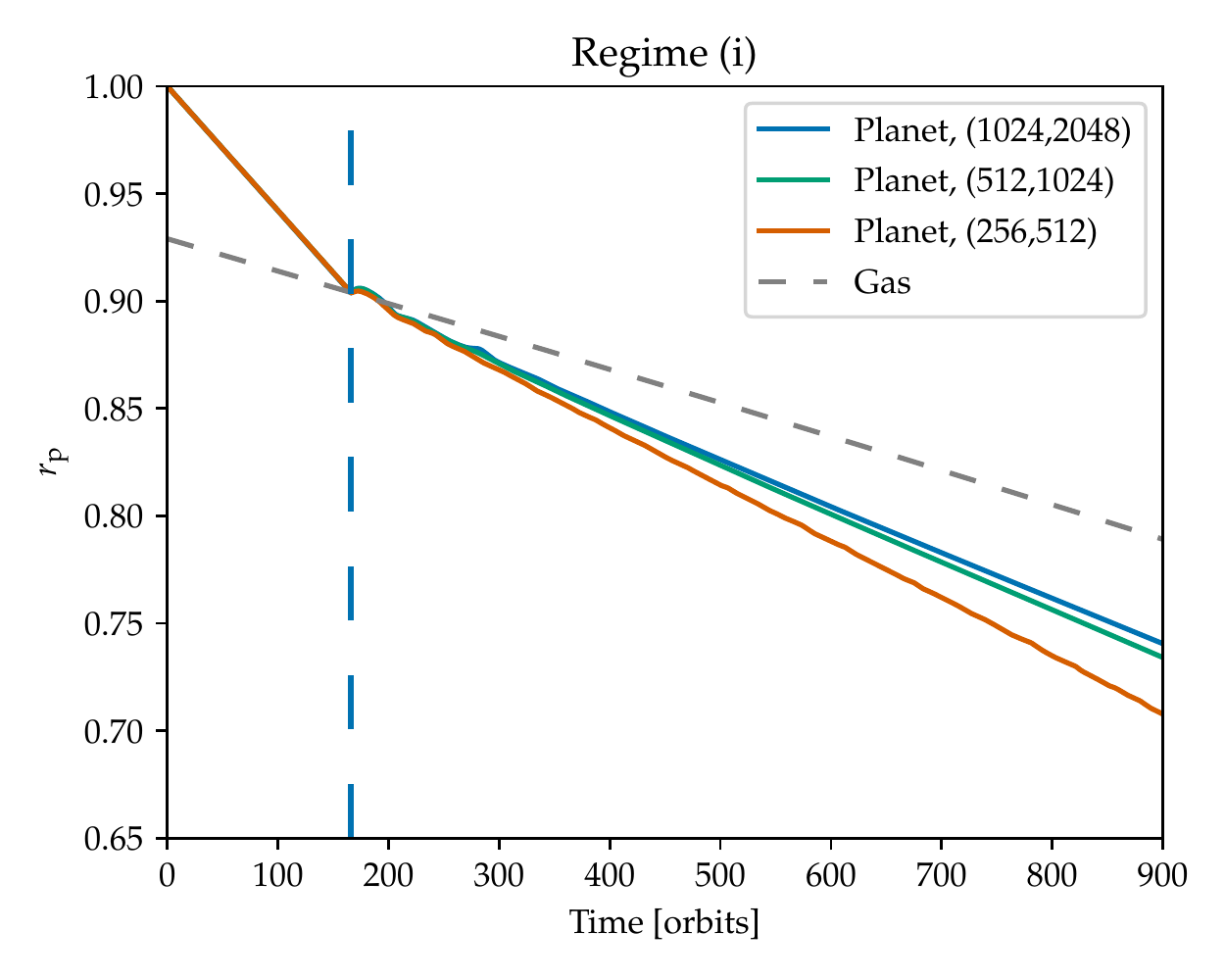}
\caption{
Experiment with forced initial migration resulting in regime (i) inward migration.
The vertical dashed blue line indicates when the planet is released. Time is given in orbits at $r=1$.
}
\label{fig:ls29}
\end{figure}

\begin{figure}
\includegraphics[width=\columnwidth]{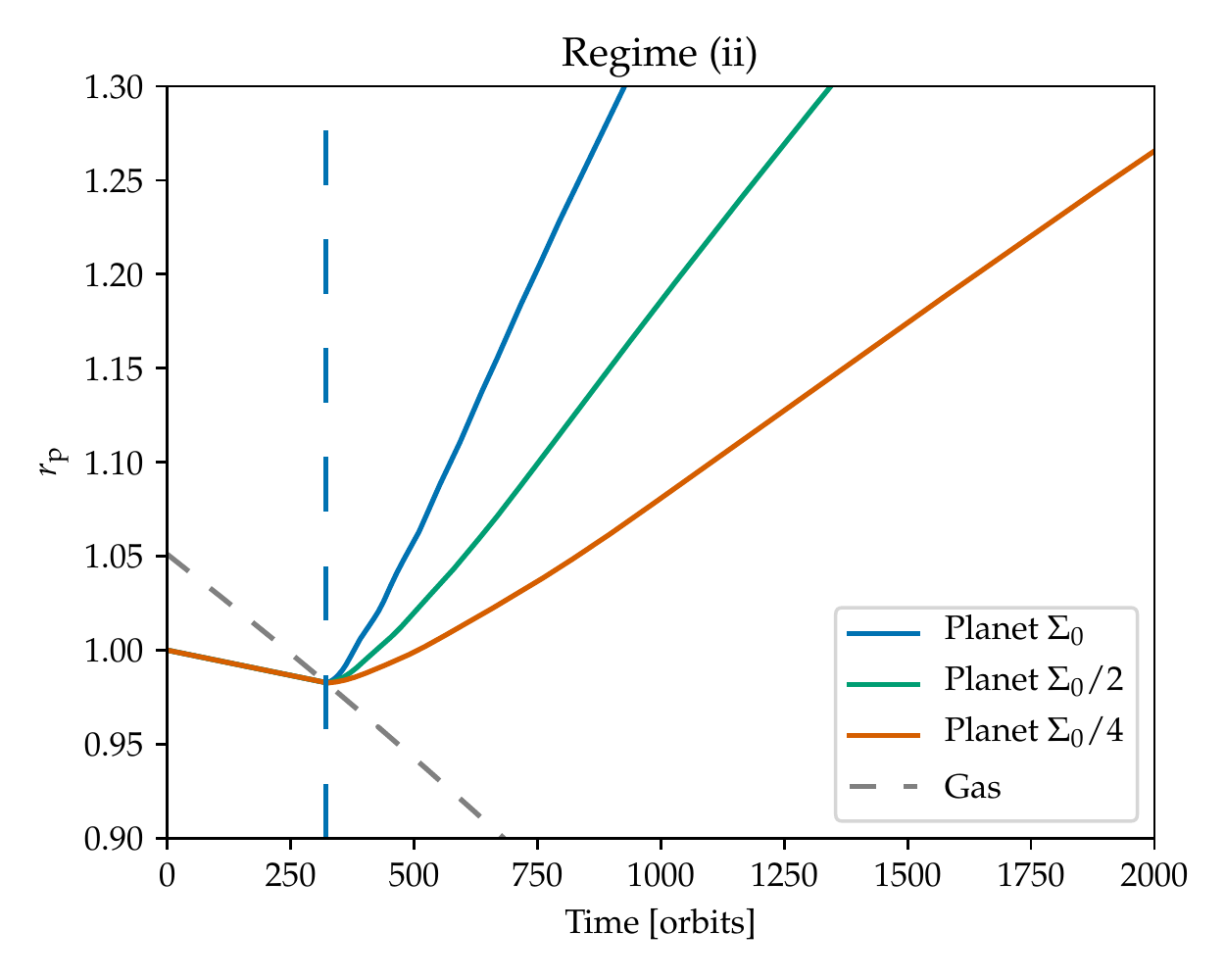}
\caption{
Experiment with forced initial migration resulting in regime (ii) runaway outward migration.
The vertical dashed blue line indicates when the planet is released. Time is given in orbits at $r=1$.
}
\label{fig:ls252627}
\end{figure}

\begin{figure}
\includegraphics[width=\columnwidth]{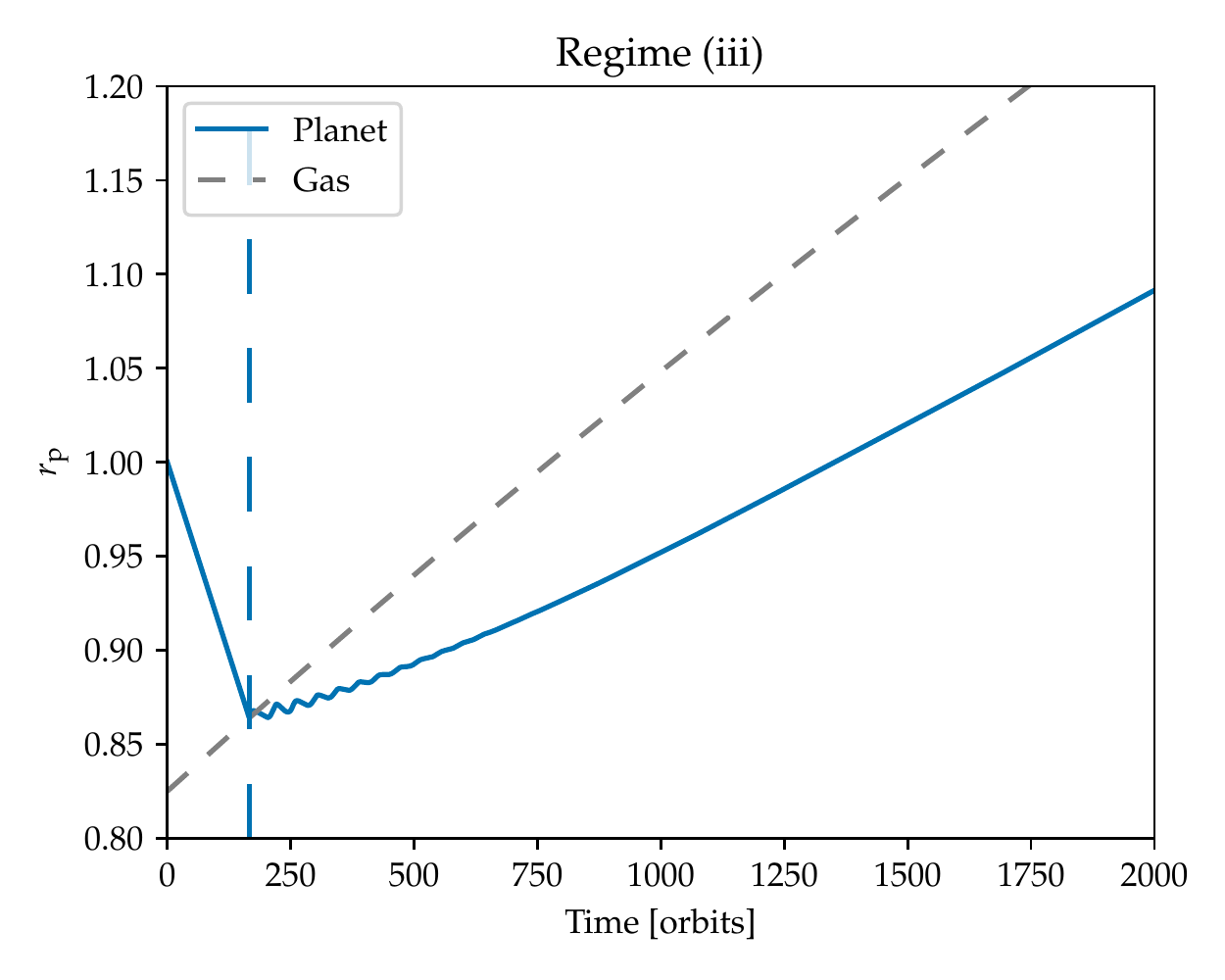}
\caption{
Experiment with forced initial migration resulting in regime (iii) outward migration.
The vertical dashed blue line indicates when the planet is released. Time is given in orbits at $r=1$.
}
\label{fig:ls30}
\end{figure}

\begin{figure}
\includegraphics[width=\columnwidth]{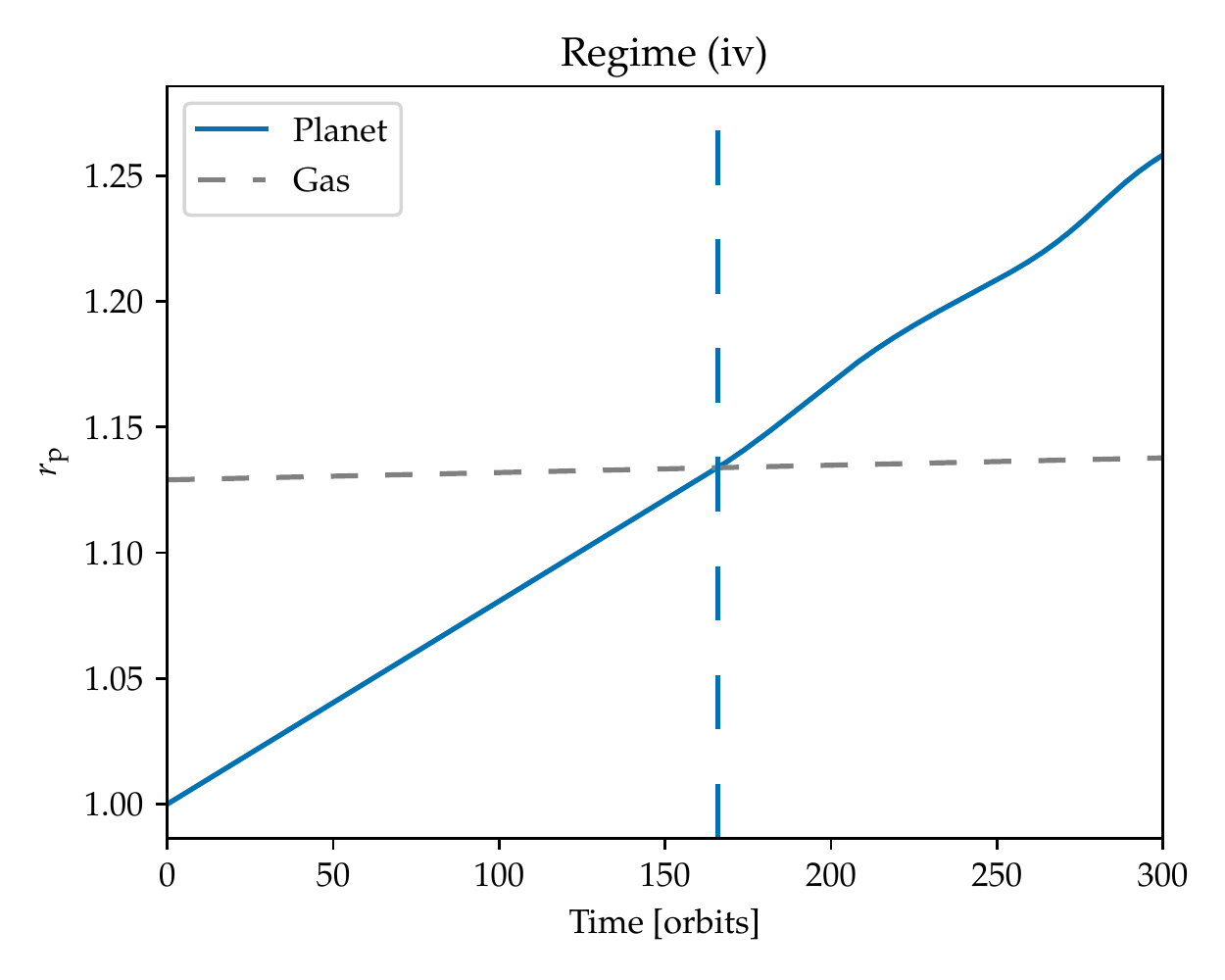}
\caption{
Experiment with forced initial migration resulting in regime (iv) outward migration.
The vertical dashed blue line indicates when the planet is released. Time is given in orbits at $r=1$.
}
\label{fig:ls31}
\end{figure}

The first experiment which demonstrates regime (i) is given in Figure~\ref{fig:ls29}. 
In this model the disk has surface density $\Sigma_0=6.3662\times10^{-3}$ and $\Sigma = \Sigma_0 (r/r_0)^{-1/2}$.
The disk surface density at $r_0=1 $ is such that the coorbital region mass ratio
\begin{align}
q_d = \frac{\pi r_0^2 \Sigma_0}{M_\star}
\end{align}
is $q_d= 1\times 10^{-2}$ and the planet mass ratio $q=\Mp/M_\star=10^{-5}$.
The body force drives an inflow with speed $\chi=3$.
For this experiment the parameters must be chosen so that the Lindblad torque driven migration rate is faster than the gas inflow rate, otherwise the planet will eventually enter regime (ii) migration.
The planet potential is turned on after $10$ orbits, and the migration from $r=r_0$ is forced until $166$ orbits, which is $2\tau_{\rm lib}$, 
and the planet is released at $4$ times the inwards radial offset that a gas parcel also starting at $r=r_0$ would have.
In accordance with expectations from equation~(\ref{eq:unified}) the planet, having started migrating inwards faster than the gas, 
slows down under the influence of the dynamical corotation torque when it is released, but is trapped migrating inwards faster than the gas.
How closely to the ideal, inviscid prediction of equation~(\ref{eq:unified}) the planet trajectory falls depends on the 
numerical diffusion induced by the finite grid resolution of the simulation.
In addition to the fiducial resolution of  $(N_r,N_\phi) = (1024,2048)$, simulations at one half and one quarter of this resolution are shown, 
and the effect of the numerical diffusion limiting the vortensity contrast in the corotation region, and hence the size of the corotation torque can be seen.
As the resolution is lowered, the planet migration rate diverges from the gas inflow rate in accordance with the expected effects of numerical diffusion
mixing the libration region vortensity with the background disc.

\begin{table}
\caption{Fast migration rates in regime (ii)}
\label{tab:fastrates}
\begin{center}
\begin{tabular}{cc}
Surface Density & $d \rp/dt$\\
\hline
$\Sigma_0$ & $9.6\times10^{-5}$\\
$\Sigma_0/2$ & $5.5\times10^{-5}$\\
$\Sigma_0/4$ & $3.0\times10^{-5}$\\
\hline
\end{tabular}
\end{center}
\end{table}

The second experiment demonstrates regime (ii), and the results are given in Figure~\ref{fig:ls252627}. 
Here the disk has a flat profile $\Sigma = \Sigma_0 (r/r_0)^{0}$ and  $\Sigma_0=3.1831\times10^{-3}$, and the inflow is set to $\chi=2$,
such that the radial gas flow is on the order of $v_r=-3.4\times10^{-5}$.
The planet potential is turned on after $10$ orbits, and the migration from $r=r_0$ is forced until $322$ orbits, which is $4\tau_{\rm lib}$.
The planet is released at $25\%$ of the inwards radial offset that a gas parcel also starting at $r=r_0$ would have.
In accordance with expectations from equation~(\ref{eq:unified}) the planet, having started migrating inwards slower than the gas, 
experiences an increasing dynamical corotation torque after it is released, and this eventually turns the planet around and causes it to migrate outwards as a runaway.
At late times the planet migrates outwards at a roughly constant rate, dependent on the surface density of the disc.
Measured outward migration rates for the three surface density values are given in Table~\ref{tab:fastrates}, 
taken as the average over the radial interval $r=[1.1,1.2]$ where the motion is roughly constant in all cases. 
Sustained fast outward migration rates are observed to scale with surface density, but the proportionality is sub-linear. 
This suggests complicated dependencies apply, and the physics which determine regime~(ii) fast migration rates is discussed in detail
in Section~\ref{sec:fastmigration}.

The third experiment demonstrates regime (iii) and the results are given in Figure~\ref{fig:ls30}. 
Regime~(iii) is the outward mode of migration with the planet asymptotically locked to the disc flow, a counterpart to regime~(i).
This disc has $\Sigma = \Sigma_0 (r/r_0)^{-1/2}$  with $\Sigma_0=3.1831\times10^{-3}$, but the radial flow is now outwards with $\chi=-2$.
The planet potential is turned on after $10$ orbits, and the migration from $r=0$ is forced until $166$ orbits, which is $2\tau_{\rm lib}$.
For regime (iii), the planet should be moving outwards more slowly than the gas, so it is initially forced inwards to $-400\%$ of the outward radial offset that a gas parcel also starting at $r=r_0$ would have.
When the planet is released, the dynamical corotation torque causes its migration to rapidly reverse, but in accordance with expectations from equation~(\ref{eq:unified}) it is limited to moving outwards somewhat more slowly then the disc gas.

The fourth experiment demonstrates regime (iv) and the results are given in Figure~\ref{fig:ls31}. 
The disc used has  $\Sigma = \Sigma_0 (r/r_0)^{0}$ and $\Sigma_0=6.3362\times10^{-3}$.
A slower radial outward flow is driven with $\chi=-5$.
The planet potential is turned on after $10$ orbits, and the migration from $r=0$ is forced until $166$ orbits, which is $2\tau_{\rm lib}$.
To produce regime (iv) behaviour the planet is released at $10$ times the outward radial offset that a gas parcel also starting at $r=r_0$ would have.
It was found that for regime (iv) migration to continue once the planet was released, that the torque on the planet at the release time needed 
to be approximately consistent with outward migration at the forced  outward migration velocity. 
Even then, the outward regime (iv) migration displays a slowly growing oscillation, possibly indicative of an overstability.

Having demonstrated that the four regimes predicted by equation~(\ref{eq:unified}) exist and can be achieved in simulations, 
we proceed to demonstrate interesting consequences in some more physically plausible situations.

\section{Reversing Migration}
\label{sec:reversingmigration}

\begin{figure*}
\includegraphics[width=\textwidth]{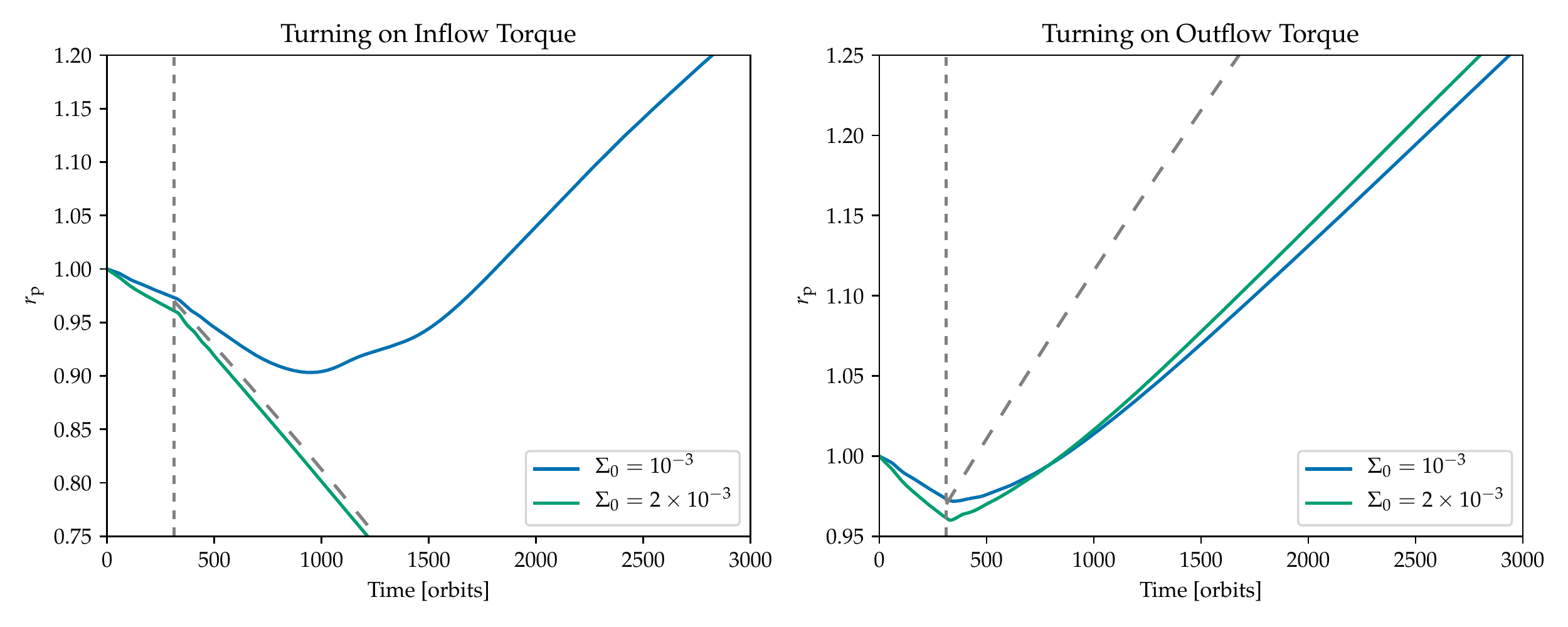}
\caption{Experiments showing the effect of introducing a disc torque on a planet migrating in a laminar disc.
{ Left panel}: Introducing a $\chi=2$ inflow torque
{ Right panel}: Introducing a $\chi=-2$ outflow torque.
In both panels the vertical dotted line shows when the disc torque begins its 10 orbit ramp up.
The dashed line shows an example gas trajectory in the laminar torqued disc.
In the left panel for an inflow torque, the fate of the planet in either regime (i) stable inward migration or regime (ii) runaway outward migration depends on the disc surface density $\Sigma$.
In the right panel for an outflow torque, the planet migrates outwards in regime (iii) in stable outward migration and cannot transition into regime (iv) runaway outward migration.
}
\label{fig:exp2b}
\end{figure*}

One interesting scenario suggested by equation~(\ref{eq:unified}) is where a planet initially migrating inwards in an inviscid disc, that is also not initially subject to an accretion-driving magnetic torque,
can experience a migration torque reversal if the protoplanetary disc's magnetic field undergoes secular evolution, and establishes a Hall-enabled laminar magnetic field torque with an associated midplane gas inflow. In other words, the planet transitions from regime (i) to regime (ii) because a magnetic torque is switched on after being absent during earlier evolution.
To model this scenario,
we let a planet migrate inwards in an untorqued, inviscid disc, and then ramp up the body force representing a magnetic field. Here it  is ramped up  over $10$ orbits,
driving gas inflow.
During the initial migration phase the corotation region acquires a vortensity deficit. Here, the planet migrates inwards in the disc from a lower to a higher vortensity region, while keeping the vortensity of librating coorbital material constant.
To make the transition and induce the planet to run away outwards in regime~(ii) migration, the vortensity of the corotation region will need to 
reverse from having a deficit to having an excess.
During this transition, the dynamical corotation torque will pass though zero as the vortensity deficit passes through zero.
As discussed 
in Section~\ref{sec:basicanal}, this point is a bifurcation in the behaviour of the system.
At this time, the migration torque will be entirely specified by the Lindblad (wave) torque.
If the disc surface density is low enough that the Lindblad torque does not drive the planet faster 
than the gas flow at this most vulnerable point in the transition, then the planet can enter regime (ii) and experience a runaway corotation torque, eventually resulting in sustained fast outward migration.
If, however, the surface density is high enough such that the inward migration driven by the Lindblad torque alone is faster than the disc inflow,
then the planet will be trapped in regime (i) and will be driven inwards, faster than the gas. 
Thus, there is a bifurcation at a critical surface density, 
below which regime (ii) migration should result, and above which regime (i) migration should occur at a speed close to the gas inflow speed. Here we demonstrate this transition via customised numerical experiments.

We tune the disc parameters with the use of the torque formula from \citet{2011MNRAS.410..293P}
to predict the Lindblad torque driven migration rate for a given planet mass and disc surface density,
and choose two values on either side of the critical surface density.
The disk has surface density  $\Sigma = \Sigma_0 (r/r_0)^{-1/2}$ with a radial flow $\chi= 2$, and a planet with mass
$q=10^{-5}$ is used.
These experiments are shown in the left panel of Figure~\ref{fig:exp2b}.
When  $\Sigma_0=10^{-3}$, below the critical surface density, the planet initially in regime (i) inward migration transitions
into regime (ii) outward migration, and eventually ends up migrating outwards at a sustained fast speed.
However, with double the surface density, that is $\Sigma_0=2\times 10^{-3}$, the planet is unable to transition into regime (ii) migration and is driven inwards with the disc gas in regime (i) migration.

In the first case, when the magnetic inflow torque is turned on, the vortensity of the corotation region is slowly reversed before runaway occurs.
The characteristic timescale for this to occur can be expressed by use of a simple model for the 
driving of the inverse vortensity of the corotation region as: 
\begin{align}
\frac{d \wc}{d t} = -\frac{d \rp}{dt} \left(\frac{3}{2}-\alpha \right)\frac{\wwp}{\rp} -(-v_r)\left(\frac{3}{2}-\alpha\right) \frac{w^2}{\wwp \rp}\ .
\end{align}
To lowest order in $w_c/\wwp$ this gives the timescale $\tau_{\rm reverse}$  to reverse a inverse vortensity contrast of $\Delta w$
\begin{align}
\tau_{\rm reverse} \sim \left\{ \left[ \frac{d\rp}{dt} - v_r\right]\left(\alpha-\frac{3}{2}\right)\right\} ^{-1} \rp \frac{\Delta w}{\wwp}\ .
\end{align}
We further discuss the sustained fast migration which occurs after this runaway phase, and conditions that can terminate fast migration, in Section~\ref{sec:fastmigration}.

In the right panel of Figure~\ref{fig:exp2b}, 
we also demonstrate that the introduction of an outflow torque with $\chi= -2$ to the disc has completely different results from the use of an inflow torque.
 If an outflow torque is introduced, both of these surface density choices result in the planet smoothly adopting regime~(iii) outward migration,
 as the vortensity contrast of the corotation region does not need to reverse for the planet to move from regime (i) to (iii).
In this regime (iii) simulation,  the planet in the higher surface density disc migrates closer to the disc outflow speed, 
as is expected from the effects of numerical diffusion in the computation.

\section{Viscous versus Laminar Accretion Flows}
\label{sec:viscous}

\begin{figure}
\includegraphics[width=\columnwidth]{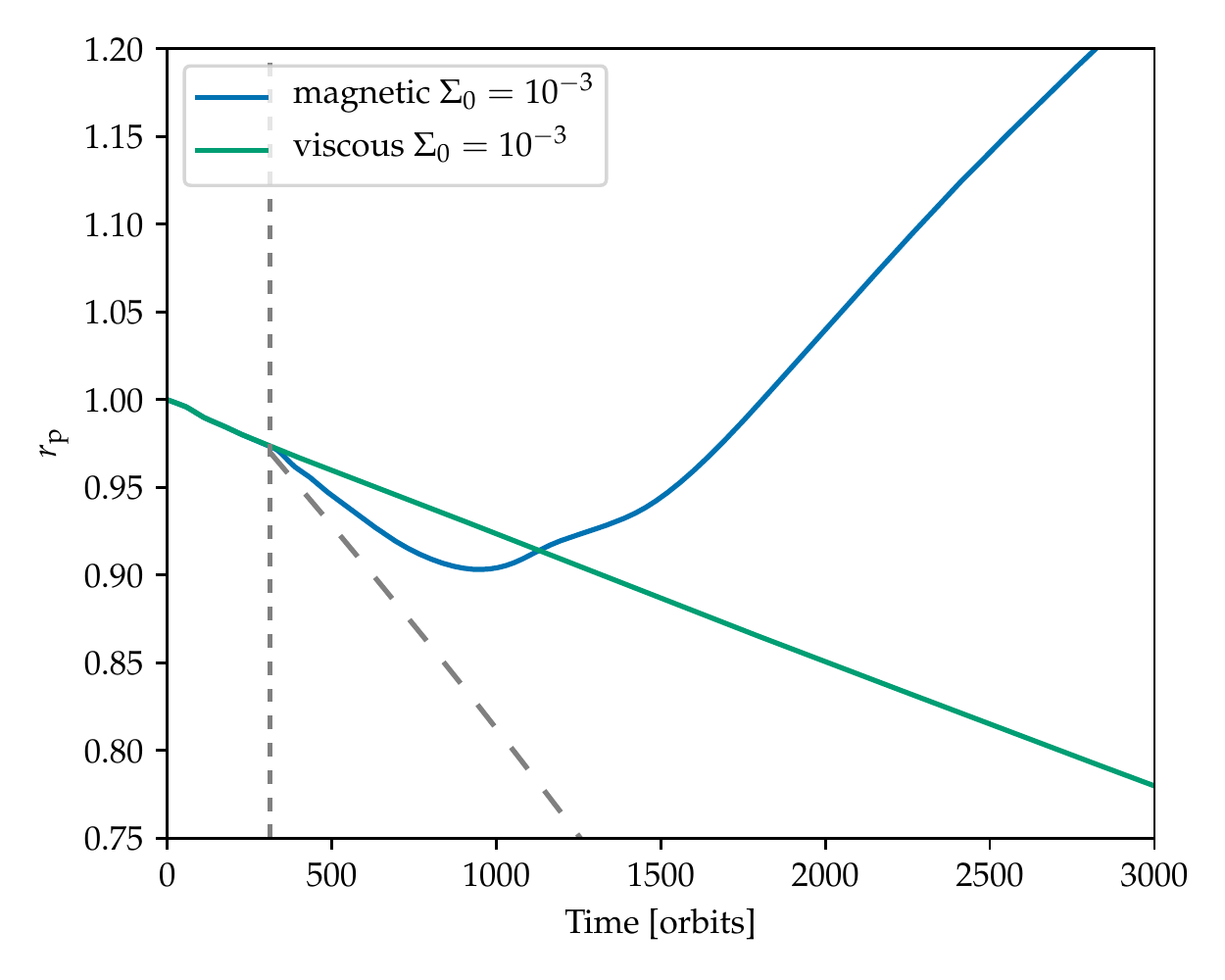}
\caption{Contrasting a laminar disc with a magnetically driven inflow, and a viscously driven inflow with the same radial velocity.
The case with an inviscid disc and a magnetic torque is repeated from Figure~\ref{fig:exp2b}, left panel, for reference.
In the viscous case, instead of reversing the migration direction due to a dynamical corotation torque, the corotation torque becomes viscously unsaturated and the planet migrates steadily inwards.
The vertical dotted line shows when the disc torque (either magnetic or viscous) begins its 10 orbit ramp up.
}
\label{fig:viscequiv}
\end{figure}

\begin{figure*}
\includegraphics[width=\textwidth]{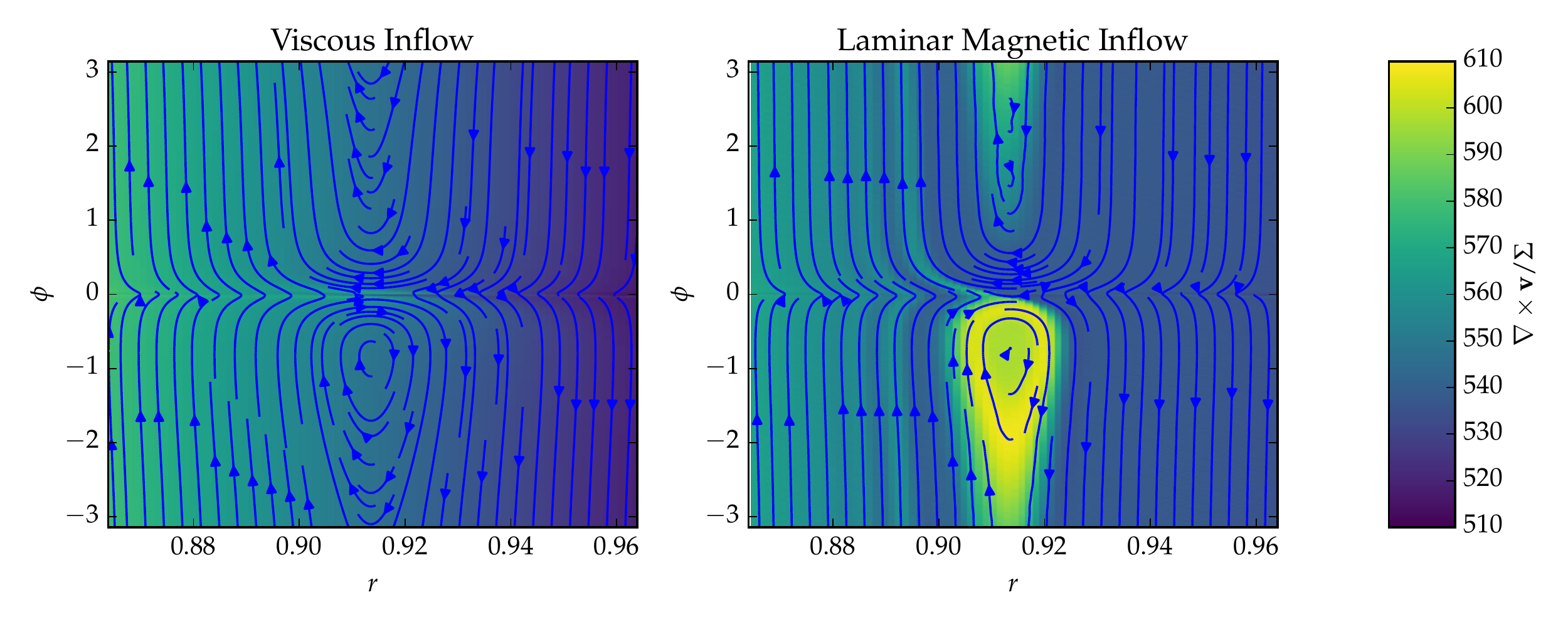}
\caption{Comparison of the coorbital region at $t=1130$ orbits in the experiment of Figure~\ref{fig:viscequiv}, at which time the two planets are at the same radial position, but migrating in opposite radial directions.
{ Left panel}: Viscously driven inflow case, the vortensity gradient across the corotation region is maintained close to the background disc value by the viscosity, resulting in an unsaturated corotation torque.
{ Right panel}: Laminar magnetic stress driven inflow case, where the vortensity of the librating streamlines evolves due to a combination of the planet radial motion and the history of magnetic torques acting on the librating material.
}
\label{fig:viscvsmag}
\end{figure*}

Given the contrasting behaviours that can be induced in the migration of a low mass 
planet by the introduction of a laminar magnetic disc torque,
it is worthwhile to contrast these to the effects of a viscous torque driving exactly the same accretion rate.
In this section, we repeat the experiment in the previous section that resulted in the planet reversing its migration,
but instead of adding a laminar magnetic field torque to the disc 
we ramp up the disc viscosity to drive an equivalent inward accretion flow.
The difference in the resulting migration of the planet serves to illustrate how the migration effects due to a laminar magnetic torque are 
\emph{not equivalent} to those generated by a viscous accretion stress.

The experiment of Section~\ref{sec:reversingmigration} is repeated here with a ramped up viscosity $\nu=\nu_0 (r/r_0)^{1/2}$ with $\nu_0=2.3\times 10^{-5}$. This produces a steady-state accretion flow that can be characterised as having an inflow at $\chi=2$.
The resulting planet migration trajectory is shown along with the equivalent result with a magnetic torque in Figure~\ref{fig:viscequiv}.
In the viscous accretion disc case, the planet continues to migrate inwards after the viscous inflow is established.
This is the opposite of what happens in the magnetically driven inflow case. 
Why this happens is illustrated in the vortensity maps shown in Figure~\ref{fig:viscvsmag}.
The vortensity in the region near the planet 
is shown at time $t=1130$ orbits, where in both cases the planet is at the same radial position, 
but is moving inwards in the viscous case and outwards in the magnetically torqued disc case.
In the viscous case, the corotation torque is fully unsaturated by the viscous diffusion of vortensity, and this is able to 
maintain the background vortensity profile across the librating streamlines in the corotation region \citep{2011MNRAS.410..293P,2014MNRAS.444.2031P}. Hence, in this case the corotation torque can be thought of as a \emph{static} corotation torque which does not depend on this history of the planet's migration, or on the history of the torques that have been applied to the corotation region, or on the planet's migration rate. It operates in addition to the Lindblad torque, and steady inwards migration results. In the inviscid case, with a magnetic torque acting on the disc,
the librating streamlines have a strongly enhanced vortensity and $\phi$-asymmetrical shape that leads to a strong \emph{dynamical} corotation torque that does depend on the history of the disc-planet system, at least until the moment when fast migration at an asymptotically constant rate sets in at late times \citepalias{2017MNRAS.472.1565M}. This illustrates why the nature of corotation torques in viscous and laminar torqued discs are fundamentally different.

\section{Fast migration}
\label{sec:fastmigration}

\begin{figure}
\includegraphics[width=\columnwidth]{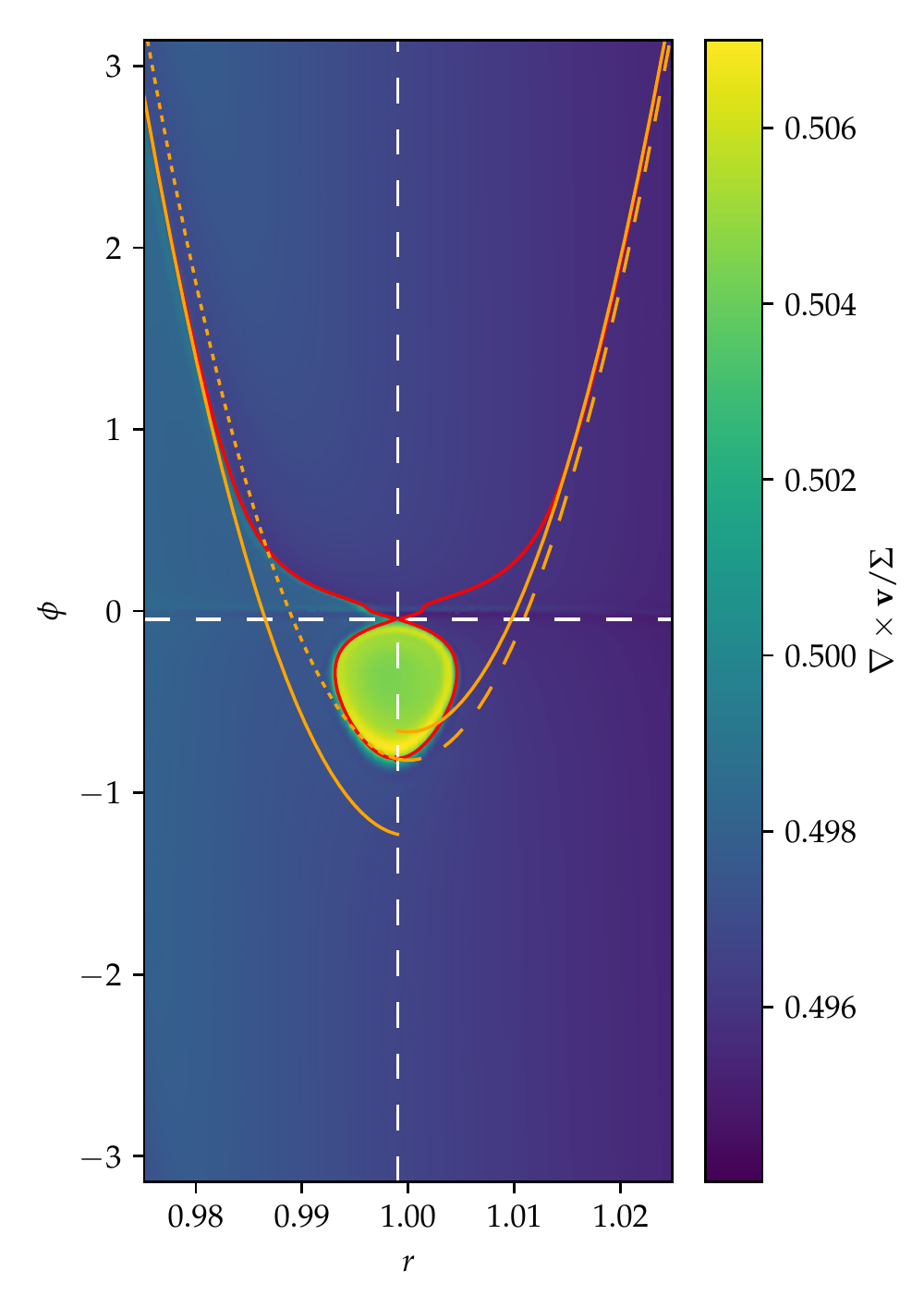}
\caption{
Analysis of $x_s$ from a simulation with $\chi=0.3$, in a disk with a very 
shallow vortensity gradient $\Sigma \propto r^{-1.4}$ after $200$ orbits.
The red streamline is the innermost one that passes behind the libration region. All other streamlines that lie interior to this one, originating in the positive $\phi$ domain, pass in front of the planet. Hence,  the red streamline effectively defines the outermost flow-through streamline. The dashed white axes show the azimuthal and radial positions of the stagnation point from which the corotation region width is measured. Note that pressure effects move this position away from the planet slightly. 
The four orange curves are streamlines in the unperturbed flow, with the planet's gravity switched off. The solid ones originate at the positive-$\phi$ sections of the red streamline.  The short dashed orange line (interior to the planet) and long dashed orange  line (exterior to the planet) originate from the tail of the libration region, which reasonable fills the space between these dashed lines.  Averaging their radial separation from the stagnation point when they cross its azimuthal location yields an estimate of $x_s =0.011$, approximately the same as the value adopted for slow migration of $0.012$.
}
\label{fig:sf85hr8_streamlines}
\end{figure}
  
\begin{figure*}
\includegraphics[width=\columnwidth]{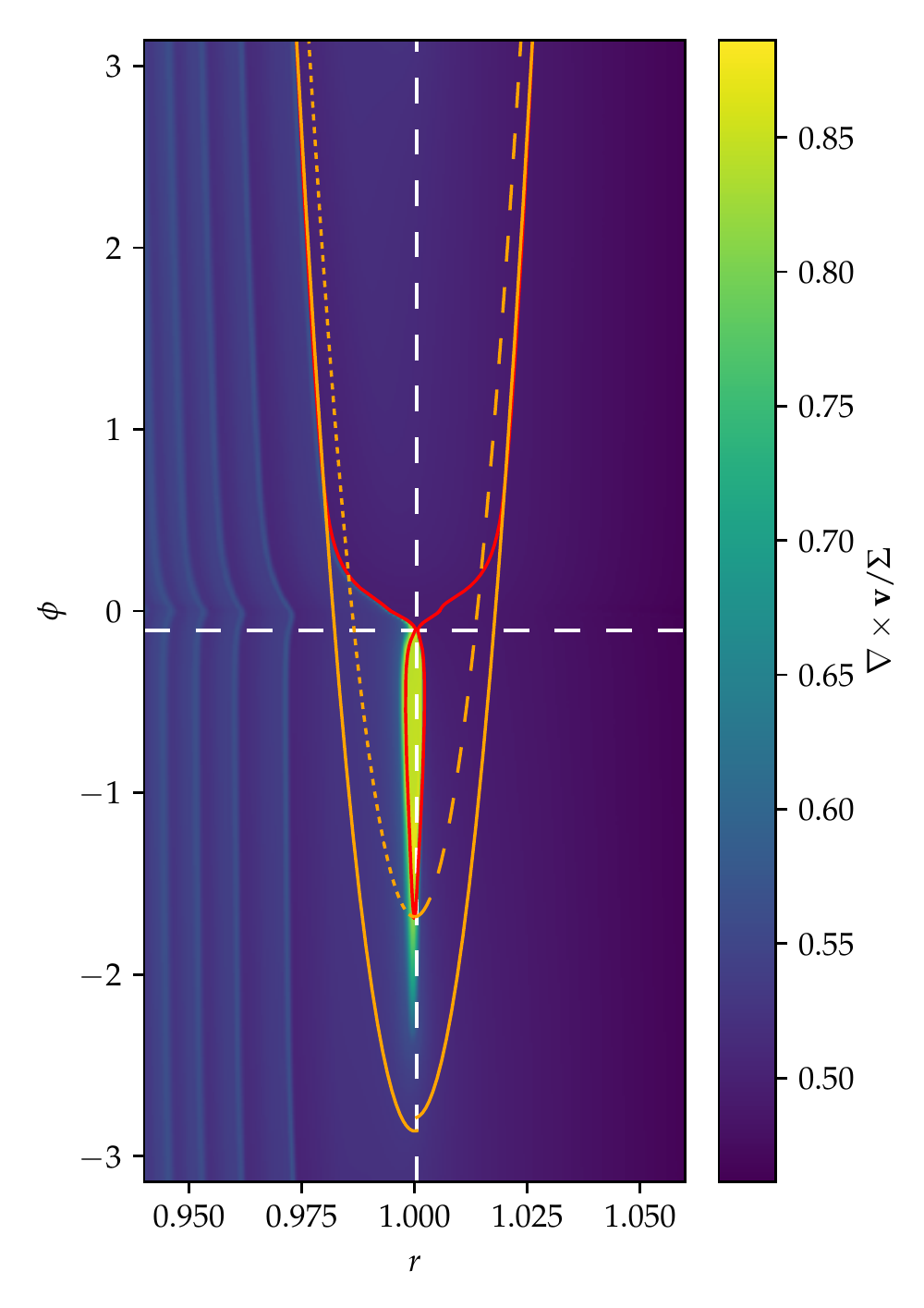}
\includegraphics[width=\columnwidth]{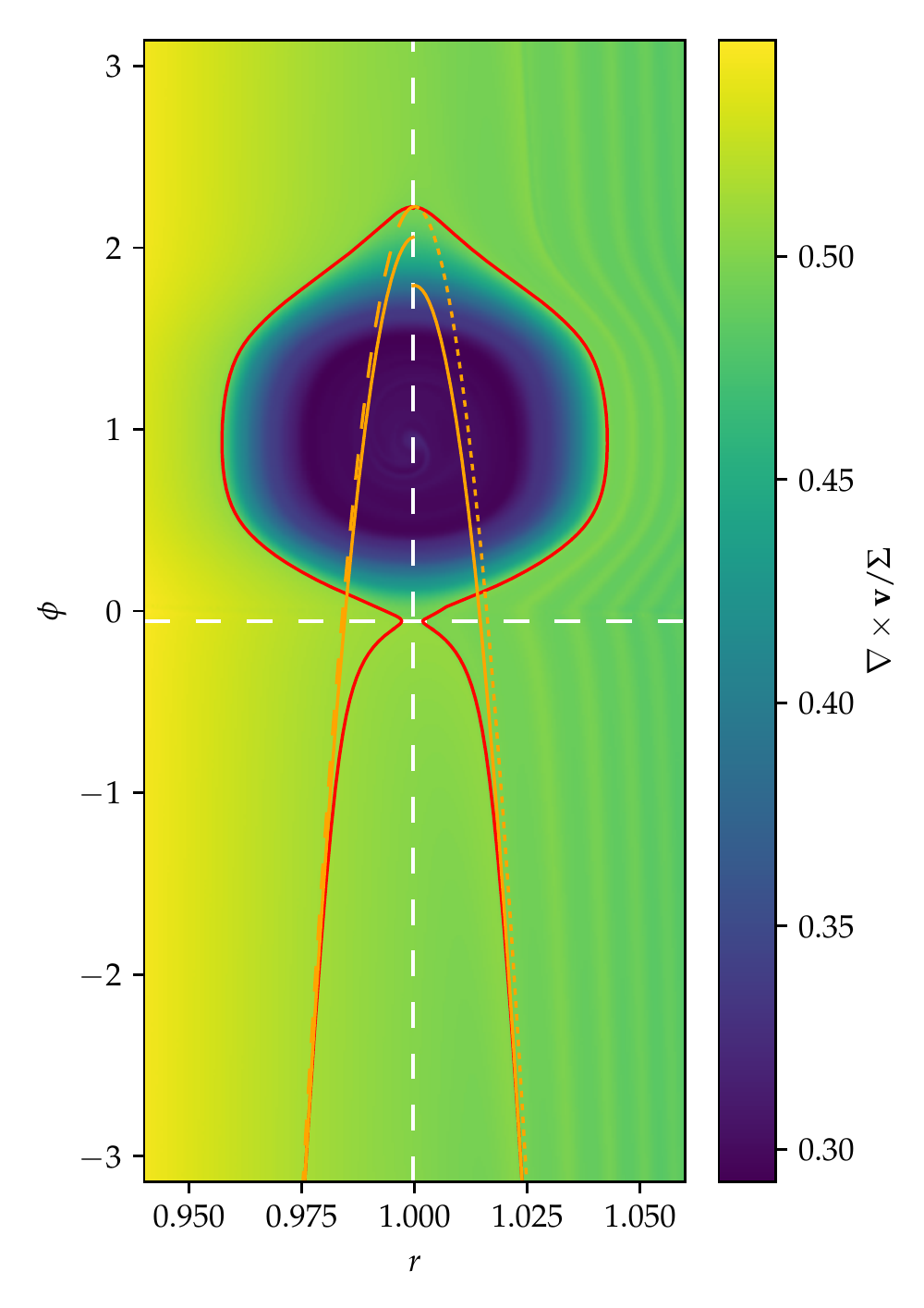}\\
\caption{
Late time flow for for $q=10^{-5}$,  $\alpha=0$, $\chi_{\rm G}=0.8$ (left panel), and  $\chi_{\rm G}=-0.8$ (right panel) both after $800$ orbits.
In the left panel, the torque is saturated and the libration island has narrowed significantly, with the tail no longer filling the space between the dashed unperturbed flow trajectories,
 particularly in relation to Figure~\ref{fig:sf85hr8_streamlines}.
In the right panel, the torque has not reached a steady state, and the libration island sits in front of the planet because of the reversed direction of the torque driving the disc flow. The island has also widened in accordance with the vortensity contrast being negative. Note that the disc aspect ratio is $h=0.05$.
}
\label{fig:scanfastex_streamlines}
\end{figure*}

\begin{figure}
\includegraphics[width=\columnwidth]{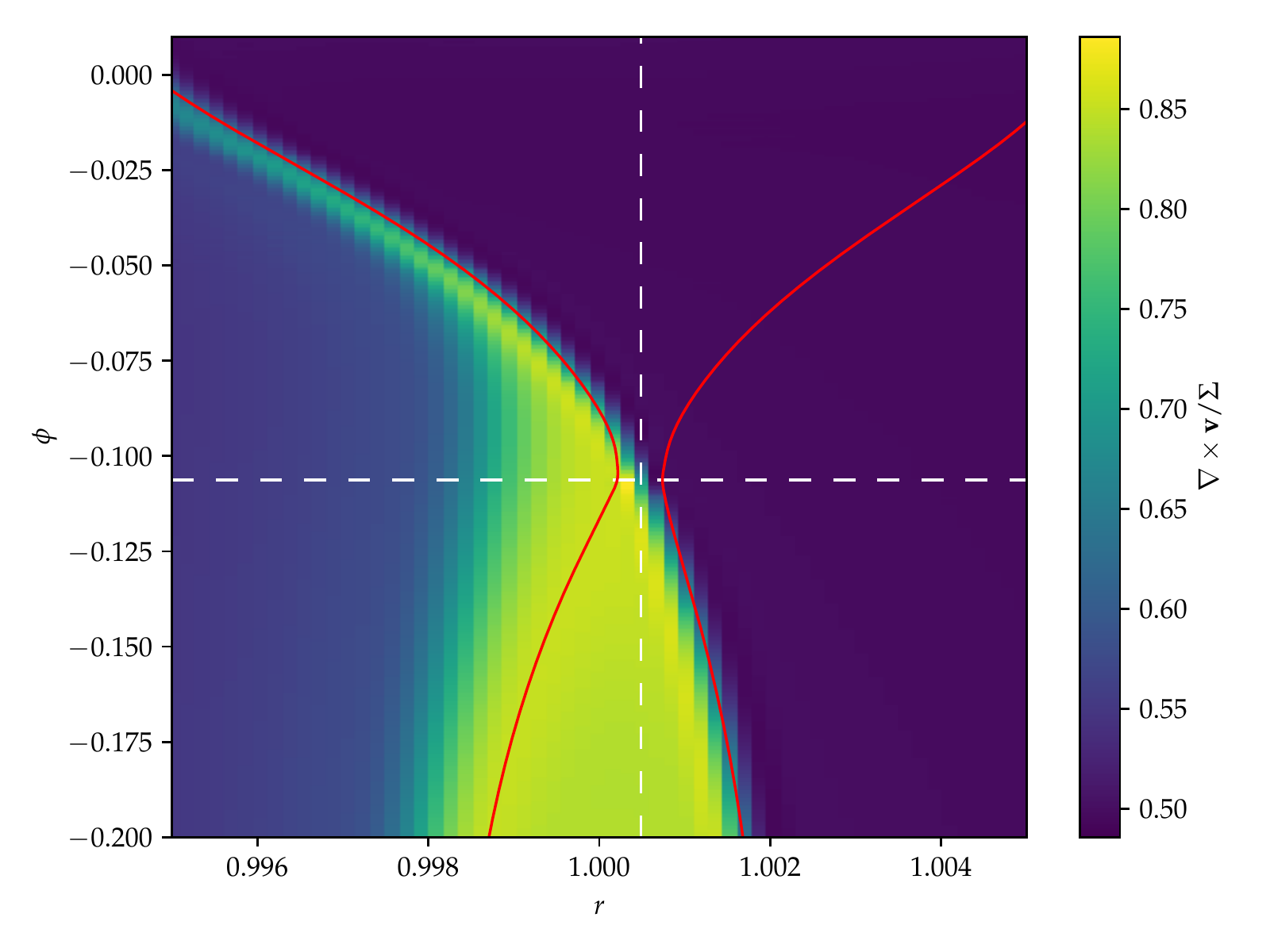}
\caption{
Zoomed section of Figure~\ref{fig:scanfastex_streamlines}, 
showing the stripping of high vortensity material off the libration island (yellow-green stream to the upper left).
}
\label{fig:scanfastex_streamlines_zoom}
\end{figure}

In Sections~\ref{sec:specifiedmigration} and~\ref{sec:reversingmigration}
experiments were shown where a planet in regime~(ii) reversed its inward migration, underwent a period of runaway outward migration, during which the migration speed accelerated, and then settled into a period of sustained migration at constant speed. The runaway phase is simply understood as outlined when this regime was predicted in \citetalias{2017MNRAS.472.1565M}:
the magnetic torque causes $\wc$ to decrease with respect to $\wwp$, and as the relative disc-planet velocity
 increases the torque and the rate of change of $\wc$ increases. However, this does not predict the existence of a sustained and stable outward migration velocity.
 
Sustained fast migration is possible when the torque stops increasing as the planet moves further and faster.
From the phenomenology of slow migration, it is reasonable to expect that a process limiting the vortensity contrast between the libration region and the disc is important in setting this rate, and as such numerical diffusion has an important role in setting the fast migration rates observed in Sections~\ref{sec:specifiedmigration} and~\ref{sec:reversingmigration}.
Thus, we instead explore this flow regime with simulations of static planets and an imposed body force driving the disk past the planet, which greatly reduces the numerical diffusion across the corotation region since this no longer migrates across the computational grid. 

For predicting the torque in the fast migration regime, we find 
that for parameters where the vortensity gradient in the disc is small, an extension of the
analytical model used for slow migration in \citetalias{2017MNRAS.472.1565M} performs well at early times.
This establishes that the key parameter in setting the torque is still the vortensity contrast between the libration region and surrounding disc.
However, for the late-time torque, which gives steady fast outward migration, this simple 
model breaks down in several ways, and it does not itself predict a limit on the (inverse) vortensity contrast in an inviscid disc. High resolution static planet simulations do, however, reveal a saturation mechanism, as described below. Having reached an understanding of why runaway migration leads eventually to steady fast migration, we decided that the most productive way to treat the resulting, somewhat complicated situation, is to seek a fit for the steady torque in the fast regime from a suite of numerical simulations. We have produced such a fit, and this is also presented and discussed below.

Before presenting the models that indicate what the saturated torque and migration rates should be in the steady fast regime, we first present representative simulations to elucidate the features of the flow in this case.

\subsection{Streamlines, angular momentum exchange and ram-pressure stripping}
For these simulation we use a static planet, as in \citetalias{2017MNRAS.472.1565M},
but also employ a radial mesh with refinement focussed on the corotation region.
The radial grid spacing is described in Appendix~\ref{sec:refinedgrid}. The use of a 
planet on a fixed circular orbit
with a fast radial disc flow is justified by the fact that the important parameter in the problem is the relative migration rate between the planet and disc, not the intrinsic migration rate of the planet. Furthermore, the phenomenon that leads to the transition from runaway migration to steady fast migration is observed in simulations with a static planet and rapid gas flow.

A first example, with $\chi=0.3$,  has a very shallow vortensity gradient, set by using $\Sigma \propto r^{-1.4}$. We remind the reader that $\chi=0.3$ corresponds to an inward radial gas flow velocity for which the time to cross the horseshoe libration region is just 30\% of the horseshoe libration period, and hence corresponds to fast migration because $\chi<1$. As we are interested in understanding what sets the corotation torque in the fast migration regime, it is worth recalling how this is determined in the slow migration limit under quasi-steady conditions before discussing how the pictures changes under fast migration.

Angular momentum is exchanged between the planet and fluid elements that undergo U-turns. In a simple symmetric switch model, applied to a disc that is flowing inwards past the planet, a fluid element that orbits outside of the planet's orbit, at a distance $\le x_s$ (defined by equation~\ref{eqn:x_s} in the slow migration limit), is forced by the planet's gravity to undergo a U-turn in front of the planet, and it jumps symmetrically from an exterior orbit to one interior to the planet. Clearly the maximum specific angular momentum exchange associated with these switches is defined by the value of $x_s$. These fluid elements are of two types. Those orbiting closer to the planet's orbit are on librating horseshoe streamlines, and their vortensity is defined by the vortensity of the corotating material, which is modified continuously by the magnetic torque that drives the radial accretion flow. Those fluid elements that orbit slightly further from the planet pass directly from the outer disc to the inner disc, and never librate. Their vortensity is defined by that in the background disc near the planet. The positive torque due to this material can be defined by summing over the inverse vortensity of all streamlines that undergo U-turns in front of the planet, as shown by equation~(\ref{eq:genhs}) below. There are also fluid elements that undergo U-turns behind the planet as gas is propelled from interior to exterior orbits, and all of these elements are trapped on librating horseshoe streamlines, such that their vortensity is defined by that in the corotation region. The negative torque due to this material can also be determined by summing the inverse vortensity of all streamlines that undergo U-turns behind the planet, and the radial width of the region that is included in this sum corresponds to $x_s$, whose value is defined as described above in the slow migration regime.

The situation is more complicated in the fast migration regime. First, the fluid elements that undergo U-turns in front of the planet all flow through directly from the outer to the inner disc. Second, the librating region and associated streamlines become confined to a ``bubble" or island that sits behind the planet, as shown in Figure~\ref{fig:sf85hr8_streamlines}, which displays results from the $\chi=0.3$ simulation described above. The fluid elements that undergo U-turns behind the planet are all contained in this bubble. Third, one needs to be careful when defining the value of $x_s$ associated with the U-turns because the fast relative flow between planet and gas causes a relatively wide region of the disc exterior to the planet to flow from the outer to the inner disc, past the orbital radius of the planet, in one synodic period, even if the planet's gravity is switched off. 

When determining the value of $x_s$ for the streamlines that undergo U-turns in front of the planet and exchange angular momentum with it, one needs to identify the streamline in the flow when the planet's gravity is switched off that corresponds to the streamline that passes closest to the planet and in front of it when its gravity is switched on. The value of $x_s$ is then the radial distance between this unperturbed streamline and the planet at the planet's azimuthal location, as illustrated by Figure~\ref{fig:sf85hr8_streamlines}. Similarly, the value of $x_s$ for the streamlines that undergo U-turns behind the planet should be determined by examining the streamline in the flow with the planet's gravity switched off that just passes through the tail of the librating bubble when the planet's gravity is switched on (because the tail of the librating bubble defines the outermost librating streamline). The value of $x_s$ corresponds to the radial distance from the planet of this unperturbed streamline when it crosses the planet's azimuthal location, as shown in Figure~\ref{fig:sf85hr8_streamlines}. Analysis of the streamlines plotted in Figure~\ref{fig:sf85hr8_streamlines} gives an estimate of $x_s=0.011$ for the fluid elements that U-turn in front of the planet \emph{and} for those that U-turn behind it, which should be compared with the value $x_s=0.012$ obtained in the slow migration regime. Hence, we see that for a disc with a small vortensity contrast between the librating island and background disc, even in the fast migration regime the value of $x_s$ closely matches that in the slow migration regime.

However, when the vortensity contrast between the libration island and the disc becomes large, the streamline geometry is modified compared to the simple picture described above.
Figure~\ref{fig:scanfastex_streamlines}, left panel, shows the vortensity map and critical streamline analysis for a
run with $\chi=0.8$ and a large background vortensity gradient set by $\Sigma \propto r^{0}$ after 800 orbits.
At this time the torque has saturated to a steady value. First, in contrast to Figure~\ref{fig:sf85hr8_streamlines}, it is clear that the libration island has narrowed significantly, so that the libration island U-turn behind the planet cannot be approximated by a symmetric switch with the same radial width as the flow through U-turn in front of the planet. This can be understood in terms of the local modification in the shear due to the change in vortensity of the libration region. As the vortensity increases, the shear in the libration region increases. The material in the libration region is thus making a horseshoe turn in a disk with a steeper rotation law that is hence narrower.
Indeed, if the vortensity perturbation in the libration region is negative, the libration island becomes wider, due to the local shear in the libration region being smaller than the Keplerian value, as shown in the right panel of Figure~\ref{fig:scanfastex_streamlines} with a negative value of $\chi_{\rm G}$.
This is in essence an extreme version of the alteration of the width of the libration region phenomenon described by 
\citet{2009ApJ...703..845C}, due in their case to the gradient of vortensity across the libration region in a viscous barotropic disc.

The above discussion does not explain why runaway outward migration is observed to level out to a constant migration rate. To understand this we note that at late times in simulations like that shown in Figure~\ref{fig:scanfastex_streamlines}, left panel, with a positive $\chi_{\rm G}$, we observe that the vortensity of the libration island has stopped growing, and a stream of 
high-vortensity material is continually stripped off from the front of the libration island and swept downstream into the disc.
This interaction is shown in detail in Figure~\ref{fig:scanfastex_streamlines_zoom}. Hence, it is the ram-pressure stripping of vortensity from the head of the librating island that explains why the runaway changes to a constant migration speed.
This ram-pressure stripping also appears to be the main source of the vortensity stripes seen downstream of the planet in Figure~\ref{fig:scanfastex_streamlines}.
That there exists a resolvable mechanism which limits the growth of the vortensity contrast of the libration island in inviscid simulations allows us to construct an approximation for the torque as a function of relative gas-planet radial motion, that in turn gives a prediction for the sustained steady-state fast outward migration speed.
Before doing so, we will develop an analytical model applicable in the low vortensity contrast regime, 
which further elucidates the nature of the fast migration regime torque beyond the discussion given above.

\subsection{Analytical model}
\label{sec:fastanalytical}
\begin{figure}
\includegraphics[width=\columnwidth]{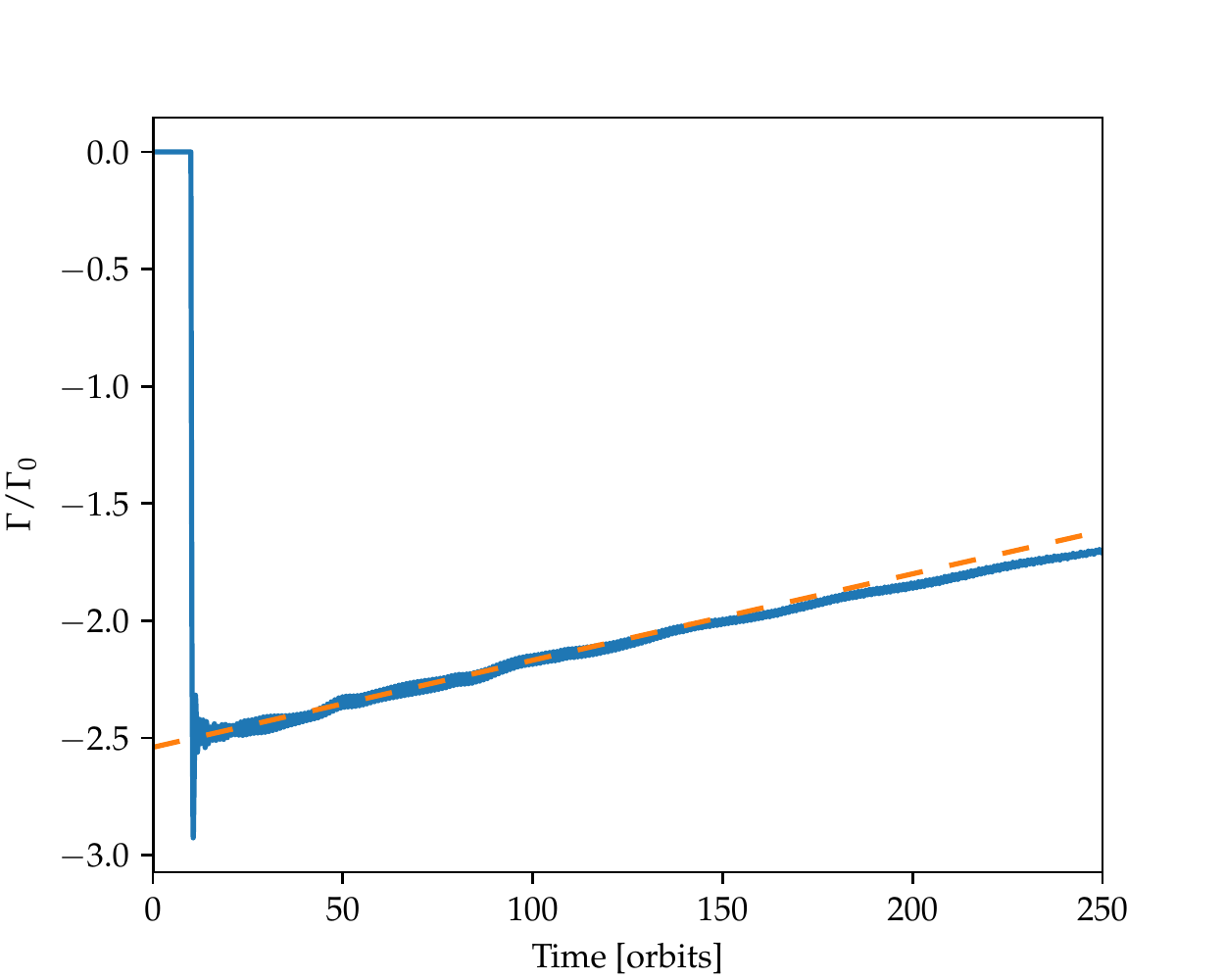}
\caption{
A $\chi=0.3$ static planet torque, with analytical model given by the dashed line.
}
\label{fig:fasttorque}
\end{figure}

Like in \citetalias{2017MNRAS.472.1565M}, we can calculate the horseshoe torque in the manner of \citet{2003ApJ...588..494M} and \citet{2014MNRAS.444.2031P}
by evaluating the integral
\begin{align}
\Gamma_{\rm hs} = \frac{3}{4} \rp \Omega_{\rm p}^3 \int_0^{x_s} \left[ w_{C_1}(x) -w_{C_2}(x)\right] x^2 dx\, , \label{eq:genhs}
\end{align}
where $w_{C_1}(x)$ and $w_{C_2}(x)$ are values of inverse vortensity on two radial cuts located just in front of and behind the two horseshoe turns, and $x$ is the radial distance between the planet and fluid streamline that makes a switch.
Formally the derivation of this expression invokes conservation of vortensity. 
To use it here we only require conservation over timescales of the U-turn, which is shorter than the libration time $\tau_{\rm U-turn} \approx h \tau_{\rm lib}$.
Thus, we apply the argument from the Appendix~A of \citetalias{2017MNRAS.472.1565M}, but to faster migration, up to $\chi \sim h$.
On the libration region turn, we know from the numerical simulations that $w_{C_2}=\wc(t)$ is constant with respect to $x$ due to phase-mixing.
The flow-through turn in front of the planet processes unmodified disc material, so to first order in $x/\rp$ the inverse vortensity 
on the integral contour drawn in front of the leading horseshoe turn is
\begin{align}
w_{C_1}(x) = \wwp \left(1+\left(\frac{3}{2}-\alpha\right) \frac{x}{\rp}\right)\ .
\end{align}
It is clear only the leading order term is needed as $w_{C_2}$ is $x$-independent, so we can write for the fast migration torque: 
\begin{align}
\Gamma_{\rm f} = \frac{3}{4} \rp \Omega_{\rm p}^3 \int_0^{x_s} \left[ \wwp -\wc(t)\right] x^2 dx\, , \label{eq:genhsfast}
\end{align}
which evaluates to
\begin{align}
\Gamma_{\rm f}
&= \frac{\rp}{4}\Omega_{\rm p}^3 x_s^3 (\wwp-\wc(t)) \ . 
\end{align}
The use of equation~(\ref{eq:genhs}) shows how the transition from the slow migration regime of \citetalias{2017MNRAS.472.1565M} 
occurs - here the two horseshoe turns become completely disconnected so no sections of the two turns cancel out.  This occurs because the turns in front of the planet consist of fluid elements that pass directly from exterior to interior orbits when the relative radial motion between the planet and disc material is fast. The turns behind the planet consist only of librating fluid elements.

In this case we analytically predict fast outward migration due to the fast torque $\Gamma_{\rm f}$ as
\begin{align}
\Gamma_{\rm f} = \frac{1}{2}\left(1-\frac{w_{\rm c}(t)}{w_0}\right)\Omega_{\rm p}^2 \Sigma_{\rm p} x_{\rm s}^3 \rp\ .
\end{align}
With a static planet, the time evolution of inverse vortensity is given by
\begin{align}
\wc(t) &= w(\rp) \left[ 1 + t/\tau_w \right]^{-1} \ ,\label{eq:wcsol}\\
\tau_w  &= \left[ \left(\frac{3}{2}-\alpha\right)\frac{( -v_r)}{ \rp} \left(\frac{\rp}{r_0}\right)^{\alpha-\frac{5}{2}} \right]^{-1}\ .
\end{align}
This form agrees well with simulations when the flow stays close to the simple model for the libration region.
For example, with a planet mass $q=5\times10^{-6}$, density power law $\alpha=1.4$ and $\chi=0.3$, 
very good agreement is found with the analytical model, as shown in Figure~\ref{fig:fasttorque}.
This model, and the excellent agreement with numerical experiment, make it clear that the dynamical 
corotation torque continues to apply even as the planet-disc relative radial velocity becomes fast ($\chi_{\rm G}<1$).
As the vortensity contrast in the libration island grows, however, we find that the flow deviates 
from the analytical model so much as to render the model inapplicable, as shown in Figure~\ref{fig:scanfastex_streamlines}. Hence, we resort to a numerical parameter study of migration torques in this area of parameter space that is presented in the next section.

\subsection{Numerical parameter study and model}
\label{sec:fastnumerical}

\begin{figure}
\includegraphics[width=\columnwidth]{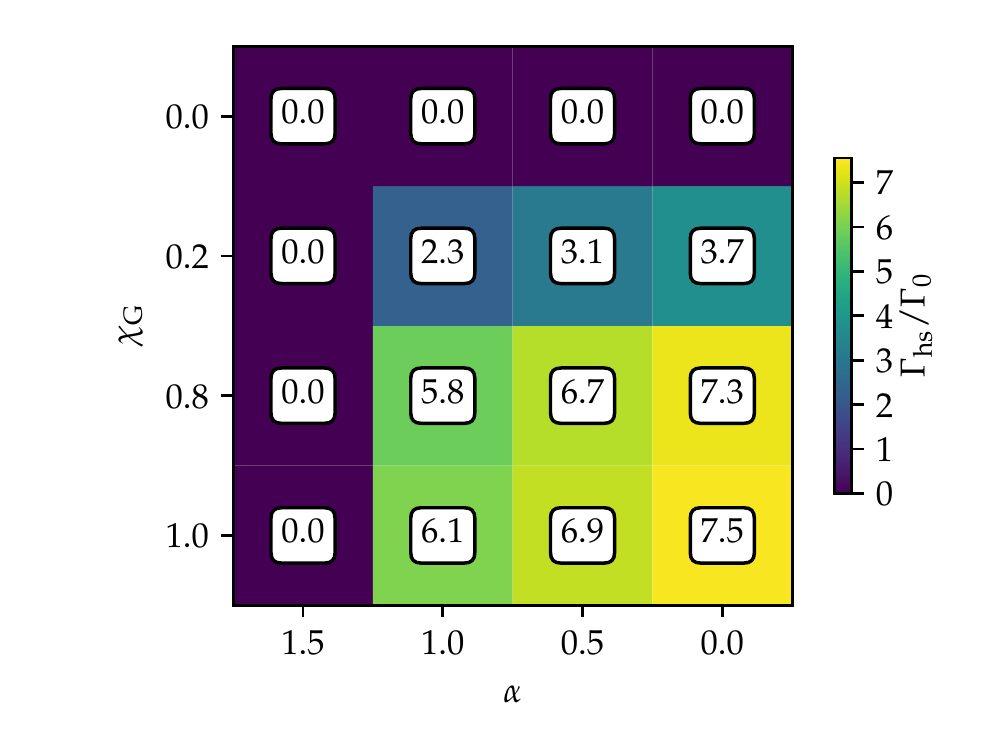}
\caption{
Values for regime (ii) saturated fast torques, as averaged over $q$ values from parameter scan simulations.
Numerical values are given for $\Gamma_{\rm hs}/\Gamma_0$ at each point in parameter space, 
and illustrated with the colour scale. 
The bordering zeros values are theoretical limits, not direct simulation values.
Note that the directions of the axes have been reversed, so that the planet-gas relative motion increases upwards, 
and the disc vortensity gradient increases to the right in the plot.
}
\label{fig:scanfast}
\end{figure}

We find in our simulations that as the vortensity contrast between the libration island and the disc grows, 
the width of  the the libration island narrows, going beyond the regime of small vortensity contrasts where the 
simple model used in the previous section applies.
Moreover, at late times, the vortensity contrast saturates due to ram-pressure stripping of high vortensity material 
off the head of the libration island, as shown in Figure~\ref{fig:scanfastex_streamlines_zoom}.
Hence, so that progress might be made in understanding the consequences of regime~(ii) migration, we present here a simple numerical analysis of the late-time torques in a inviscid disc for a range of disc and planet parameters.

To produce a model for the late-time dynamical corotation torque on a planet in regime~(ii) migration,
we have run a parameter scan of static planet simulations with varying planet mass, disc flow rate,
and surface density gradient.
Planet masses $q=10^{-5}$, $5\times 10^{-6}$, $1.25\times10^{-6}$,
disc flow rates $\chi_{\rm G}=1.0$, $0.8$, $0.2$
and surface density radial power law indices $\alpha=1.0$, $0.5$, $0.0$.
Expending on the order of a million CPU hours on evolving these simulations,
we found late-time convergence to a constant torque, although in two cases ($q=1.25\times10^{-6}$, $\alpha=0.5$, $\chi_{\rm G}=1.0,$ and $0.8$) 
we were not able to evolve the simulation for a long enough physical time.
However, for the cases which did evolve well, we found no significant dependence of the dimensionless total torque, $\Gamma/\Gamma_0$, on the planet mass ratio $q$. 
Therefore, we averaged the values  obtained for different $q$, and to derive the final measured $\Gamma_{\rm hs}$ corotation torque we 
subtracted from the total torque measured in our simulations the Lindblad torque as predicted by the formulas in \citet{2010MNRAS.401.1950P}.
We include a reasonable limit of zero corotation torque when $\alpha=-1.5$ (zero vortensity gradient) in the fit, 
although it is not part of the simulated parameter scan.
The numerical values are reported in Figure~\ref{fig:scanfast}, and in Appendix~\ref{sec:prescription} we provide a prescription for the torque as function of the surface density power-law index, $\alpha$, and the disc accretion flow parameter, $\chi_{\rm G}$, based on a simple bi-linear interpolation of the values given in Figure~\ref{fig:scanfast}. We also give a more extended discussion there about how one may practically model a magnetically torqued disc with an embedded planet in the context of N-body simulations of planet formation or population synthesis models.

\subsection{Termination of fast migration}
\label{sec:termination}

\begin{figure}
\includegraphics[width=\columnwidth]{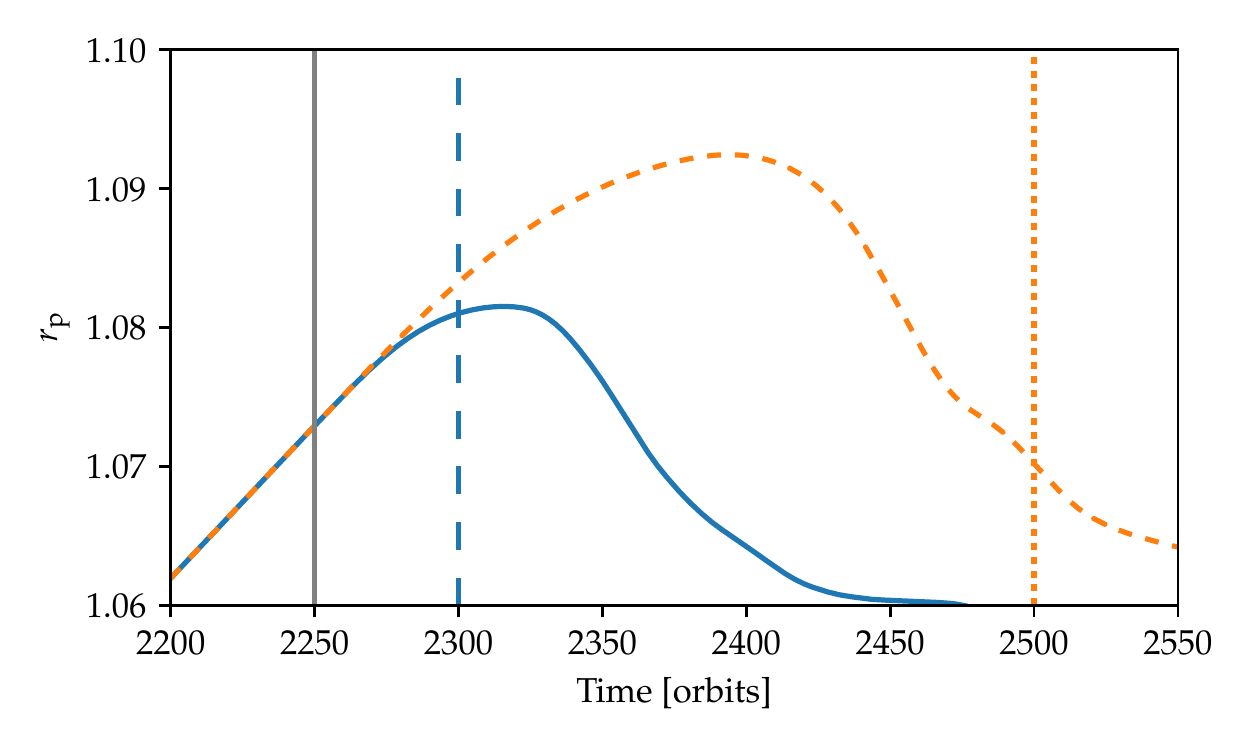}
\caption{Trajectories of a planet dropping from regime~(ii) to regime~(i) migration as the magnetically driven inflow is reduced.
Solid curve: inflow ramped down over 5 orbits from the vertical solid line to the vertical long dashed line, 
Dashed curve: inflow ramped down over 250 orbits, five times slower than in the first case from the vertical solid line to the vertical short dotted line.
Once the planet has slowed, the fast drop to regime~(i) occurs on timescales connected 
to the horseshoe turn rather than the timescale of the external forcing of the disk inflow, 
as this change is driven by the widening of the libration island.
}
\label{fig:ls45r8_drop}
\end{figure}

\begin{figure*}
\includegraphics[width=\textwidth]{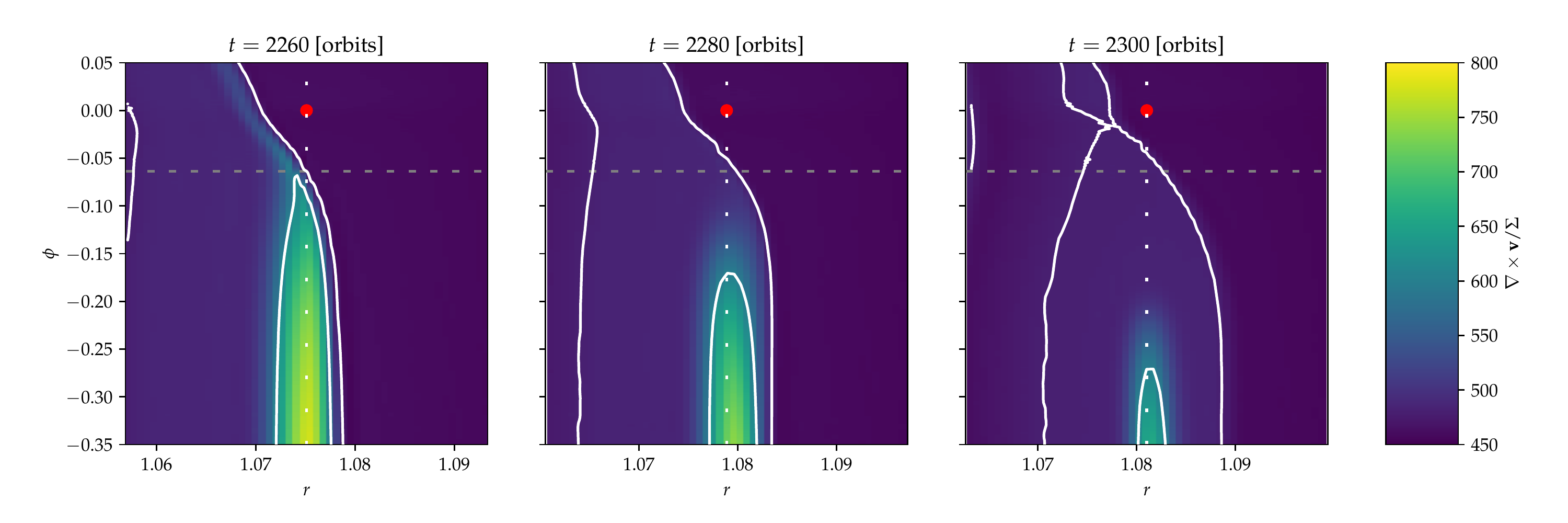}
\caption{
The change in the libration region causing the transition from regime (ii) to regime (i).
The width of the panel is $x_s=0.012$ to either side of the planet. White contours show two iso-vortensity levels.
A red dot shows the planet position, and a vertical dotted line shows the planet radius.
A dashed grey horizontal line shows where the $\nabla\times\mathbf{v}/\Sigma =480$ 
contour initially crosses the planet's radial position.
As the planet-disc relative velocity decreases, more low-vortensity material starts making the 
horseshoe turn behind the planet, and this contour shifts upwards and to the right on each panel.
}
\label{fig:dropbubble}
\end{figure*}

As the previous section has established the mechanisms which allow sustained fast regime~(ii) outward migration to occur, we explore here one mechanism by which such an episode can end.
Experiments like in Section~\ref{sec:reversingmigration} have explored the migration behaviour of a planet when the disc inflow torque is increased, so here we explore the effect of decreasing this torque with time on a planet in regime~(ii) migration. This can cause the planet to terminate its fast migration, and return to regime~(i).

Thus, our experiment is an extension in time of the run in Section~\ref{sec:reversingmigration} 
with $\Sigma_0=10^{-3}$, $q=10^{-5}$,  and an inflow torque $\chi=2$,
but now $v_r$ is ramped to zero by decreasing $B_0$ in equation~(\ref{eq:vr}) over $50$ or $250$ orbits starting at $t=2250$ orbits.
The planet trajectories for this relevant part of the simulation are shown in Figure~\ref{fig:ls45r8_drop}.
The vortensity profile as shown in Figure~\ref{fig:dropbubble} is smooth, due to numerical diffusion as the planet moves radially across the grid.
However, it is still clearly much narrower than the canonical value of $x_s=0.018$ which would be expected in the left panel, as judged by the width of the inner white contour.
As shown in Figure~\ref{fig:ls45r8_drop}, although the ramping timescale varies by a factor of five, 
the timescale for the planet to transition to inward regime~(i) migration does not vary by such a large factor.
Details of the flow shown in Figure~\ref{fig:dropbubble} offer an explanation for this.
As the planet-gas relative velocity slows, an amount of low-vortensity-contrast material that was originally outside of the libration island begins making the 
horseshoe turn behind the planet along with the preexisting libration region material. 
Although the planet is constantly moving outwards in the disc, in the sequence of three panels the outer white contour of vortensity moves steadily rightwards to the outside of the planet's position, indicating the flow of this low vortensity material around the horseshoe turn. Hence, the material making the rear horseshoe turn becomes more symmetrical with that making the front horseshoe turn, decreasing the corotation torque expected from equation~(\ref{eq:genhs}). Given this phenomenology, one expects that the timescale for the drop from regime~(ii) to regime~(i) migration should occur on timescales related to the horseshoe turn.
For the narrowed libration region observed at $t=2260$ with half width of $\sim 0.003$ the associated  $\tau_{\rm U-turn}$  is $\sim 26$ orbits, whereas for the full horseshoe turn width as applies with small vortensity contrasts of $\sim 0.018$ the associated timescale is $\sim 4.5$ orbits. The termination of fast regime~(ii) migration evidentially involves motions of gas across this range of timescales.

When planets in regime~(ii) reach a radial position such that the magnetic field driven disc inflow significantly decreases, the most simple scenario is that they will drop to regime~(i) migration and asymptotically stall with respect to the gas. This forms a basic conjecture for the fate of  planets in regime~(ii) as they reach the outer edge of the dead zone, or enter a region of the disc where the magnetic torque acting on material near the midplane undergoes a significant decrease.

\section{Viscous steady states}
\label{sec:viscstedystates}
Our discussion of corotation torques in a magnetically torqued disc to this point has focussed on inviscid discs, where
there is no mixing of vortensity between streamlines, only evolution of the vortensity on those streamlines due to forces exerted by ordered large scale magnetic fields. In principle, however, some level of turbulent diffusion may also be present in a protoplanetary disc, either because the magneto-rotational instability operates intermittently at a low level, as described by \citet{Simon2015}, or because a hydrodynamic instability, such as the vertical shear instability \citep[e.g.][]{Nelson2013}, operates in the presence of the magnetic torques.

This issue is relevant because our torque model expressed by equation~(\ref{eq:unified}) does not predict a steady-state for a live planet beyond asymptotically locking the planet to the disc radial flow (no relative radial motion between the disc gas and planet).
However, when a finite viscosity is introduced to mix the vortensity of the libration region with the surrounding disc flow, the
vortensity contrast and corotation torque can settle into a steady-state even when there is relative motion between the planet and gas. Thus, we have extended the approach taken in \citet{2014MNRAS.444.2031P} for accounting for a finite turbulent viscosity in the disc to the situation at hand. This gives rise to a steady-state corotation torque. 

Our analysis of this problem, which includes an analytic study and a suite of numerical simulations that are compared with the analysis, is presented in Appendix~\ref{sec:viscmodel}. Generally speaking, decent agreement between the two is obtained in the parameter domain for which the analytical theory applies, hence providing a means of predicting the corotation torque in a magnetically torqued disc in which diffusion of vortensity also occurs.

\section{Discussion}
\label{sec:discussion}
Our analytical calculations and simulations show a diversity of behaviour for low mass planets in laminar magnetically torqued discs, including migration that is essentially locked to the disc flow, and runaway outward migration that saturates and transitions to steady fast migration. A significant limitation in our attempts to model dynamical corotation torques in the inviscid limit has been the effect of purely numerical diffusion in mixing the vortensity of the libration island and background disc flow, 
which inherently leads to resolution-dependent migration rates as demonstrated in Figure~\ref{fig:ls29}. In regimes~(i) and (iii) we can understand from these simulations the underlying mechanisms, and extrapolate to a concept that the planet migration should be asymptotically locked to the radial motion of the disc. In the simulations of fast regime~(ii) migration in Sections~\ref{sec:specifiedmigration}, \ref{sec:reversingmigration} and~\ref{sec:termination}, however, the analysis predicts continued runaway migration whereas the simulations demonstrate saturation of this runaway. Although our adoption of a refined grid and a non-migrating planet allowed us to explore fast migration under idealised conditions with reduced influence of numerical diffusion, and resulted in a natural saturation mechanism for runaway migration being identified, it is probable that the computed migration rates are slower than would be found in a fully inviscid disc. In these latter calculations the low numerical diffusion has come at the price of not having a fully dynamic planet, and hence the radial remapping technique of \citet{2016ApJ...826...13B} would be of particular utility in this respect. Furthermore, it is noteworthy that we have adopted a very simple equation of state in this study that removes baroclinic effects, and it is possible that under more general conditions instabilities might come into play as the vortensity contrast between the corotation region and surrounding disc grows. Such an instability was identified in a disc with a radial temperature gradient by \citet{2014MNRAS.444.2031P}, and this effect might play a role in regulating the vortensity contrast and corotation torque under more realistic disc conditions.

Both inward and outward runaway migration can occur for higher mass (e.g.\ Saturn-mass) planets in viscous discs \citep{2003ApJ...588..494M}, so some discussion of the differences between that phenomenon and the runaway migration discussed in this paper are warranted.
The drift rate during regime~(ii) outward migration is fast, in the sense defined by \citet{2003ApJ...588..494M}, and is driven by corotation torques. 
Unlike in the classical analyses of so-called Type-III migration \citep{2007prpl.conf..655P,2008MNRAS.386..164P,2008MNRAS.386..179P,2008MNRAS.387.1063P}, however, vortensity-related dynamical corotation torques can have either a negative or positive feedback on planet migration, depending on the  surface density power law $\alpha$ \citep{2014MNRAS.444.2031P}, whereas Type-III migration torques have only a positive feedback on planet migration.
 For inviscid discs with surface density power laws $\alpha < 3/2$, runaway migration of low mass planets  only exists in an outward-migrating form. 
Additionally, Type-III migration requires, for a given planet mass, a disc surface density above a critical value, whereas the torque reversal and runaway of our regime~(ii) requires a disc surface density below a critical value (Section~\ref{sec:reversingmigration}). Indeed, the mechanisms by which these two migration phenomena can be triggered and terminated also appear to be different.
How fast regime~(ii) phenomenology transitions to Type-III  as the planet mass under 
consideration is increased remains as an interesting and as yet unexplored problem.

Our analysis in this paper applies to the classical Type-I migration regime in which the planet mass is small enough to not perturb the disc local surface density significantly. For an inviscid disc without an accretion flow near the midplane, however, nonlinear steeping of the spiral density waves launched at Lindblad resonances can lead to gap formation \citep{Goodman2001} for relatively low mass planets, and to the phenomenon of the \emph{inertial limit} discussed originally by \citet{Hourigan1984}, and considered more recently by \citet{Rafikov2002}, for which migration stalls completely when a planet forms a radially asymmetric gap \citep{Li2009}. The question of how things change in a laminar disc with a significant midplane accretion flow has not yet been addressed, but will almost certainly lead to a modification of previous estimates of the planet masses for which gap formation and migration stalling occur. Planets migrating in regimes~(i) and (iii), for which the planet approximately locks to the radial drift of the disc, are of particular interest because, in a frame that migrates with the planet, there is essentially no relative radial motion between the disc and planet, and hence no flow to oppose the gap forming torques. A relative flow will develop, however, if the onset of gap formation leads to a slowing of planet migration, so understanding the interplay between migration, gap formation and laminar accretion flows for low mass planets is an interesting issue for future study, and one that can address important issues in planet formation such as the pebble isolation mass \citep{Lambrechts2014}.

Having adopted a very simple equation of state, we have not considered the role of entropy-related dynamical corotation torque effects as have been discussed for viscous discs by \citet{2015MNRAS.454.2003P} and \citet{2016MNRAS.462.4130P}.
We expect that such entropy-related torques do not have the Galilean-invariance properties of the vortensity-related dynamical corotation torque because the rate that they build up is not proportional to the relative planet-disc gas radial velocity, but only the radial motion of the planet with relation to the disc's entropy structure.
However, the asymmetry of the libration island itself does have the same Galilean-invariant property.
The basic expectation would be that an entropy contrast \citep[or entropy deficit in the terminology of][]{2015MNRAS.454.2003P} 
retained by the trapped librating material will lead to a extra component of the dynamical corotation torque.
For a non-viscous disc that is passively heated by the star, we would generally expect the entropy gradient to be positive, with entropy increasing as one moves away from the star. In the quasi-adiabatic limit, with inefficient heat exchange between the corotation region and the surrounding disc, a planet migrating inwards in regime~(i) would migrate inwards at a faster rate due the influence of the entropy. A planet that is in regime~(ii), but which starts to migrate inwards before turning around and migrating outwards would display more complicated behaviour. The entropy contrast would first slow the inwards migration, and would make the initial phase of the outward runway migration faster. Once the planet passes its point of origin, however, then the entropy contrast would change sign and would act as a drag on the outward migration. 
This scenario would be modified if instead the radial entropy gradient in the disc was negative, possibly due to magnetic dissipation accompanying the laminar magnetic stress or related to the dissipation in a disc wind.
We note that, as opposed to being relaxed by viscous mixing and/or ram-pressure stripping, the entropy contrast of the libration island can be relaxed by radiative transfer effects. Unlike turbulent viscosity, this may not have a large variation between wind-driven and turbulent disc models. Clearly, the joint analysis of entropy-related and vortensity-related corotation torques in appropriate passively-heated wind-driven disc models is an additional area of interest for further work.
The new rich behaviour introduced by  the consideration of magnetically driven radial flows in laminar dead zones
can be expected to produce a multitude of new scenarios for planet formation. Some possibilities which warrant exploration include:
\begin{itemize}
\item A low mass planet at a small radius enters regime~(ii) fast outward migration, exits that regime where the disc torques change significantly, and undergoes a regime~(i) migration stalling at large radius where it is able to increase its mass by accreting locally.
\item Multiple planets in a disc evolve via regime~(i) migration, such that their drift rates are controlled by the radial gas flow, altering the expectations for convergent Type-I migration and the production of resonant pairs and chains of planets.
\item Changing disc ionisation conditions, due to the evolution of the disc surface density and dust, lead in turn to changing non-ideal MHD conditions, magnetic field structures, and hence radial flows and coupling regimes, leaving a signature on planet formation.
\item Secular changes in the magnetic field configuration of the inner disc can be expected, so the motions of low mass planets and cores changes at different points in the disc's lifetime.
\item Planets that form in discs where the vertical magnetic field is parallel to the disc rotation axis have their migration histories strongly influenced by Hall-effect-induced radial gas flows near the midplane. Planets that form in discs where the field and rotation vector are anti-parallel do not experience radial gas flows near the midplane, as Hall EMFs are ineffective, and hence can stall their migration completely due to dynamical corotation torques.
\end{itemize}
As full global non-ideal MHD simulations with sufficient dynamical range in space and time to both resolve the evolution of the disc, magnetic field, thermodynamics and chemistry, and the evolution of planets and their orbits and interactions are still computationally speaking, some ways off, we expect that the exploration of many of these scenarios will need to be conducted using the types of customised hydrodynamic simulations that we have presented in this paper, and using reduced N-body models, using parameterised versions of the migration torques that are presented in appendix \ref{sec:prescription}.

\section{Conclusions}
\label{sec:conclusions}

In \citetalias{2017MNRAS.472.1565M} we predicted that there are four regimes of migration for a low mass planet in a laminar magnetically torqued disc, and in this paper we have demonstrated that all four regimes can occur. We have demonstrated that inwardly migrating planets can undergo migration torque reversals
and migrate outwards due to the introduction of laminar magnetic disc torques. When magnetic torques drive rapid gas inflow, we have shown that for a given planet mass a critical disc surface density exists, below which torque reversal and outward runaway migration is possible. We have demonstrated that this runaway migration eventually saturates, and transitions to steady fast migration, and we have identified the process that causes this transition. Our simulations show that if the magnetic torques acting on the disc decrease over time, the planet drops out of the fast migration regime and migrates inwards at approximately the flow rate of the gas. This is a general result for slower gas flows, as we have shown that in this case planet migration can approximately lock to the drift of the gas, such that the planet migration rate is determined by the radial velocity of the gas.

Our study has also examined why the corotation torque in a disc where viscosity drives radial gas flows differs from that in a laminar magnetically torqued disc, and we have demonstrated the critical role of viscous diffusion in smoothing vortensity contrasts such that the different behaviour is established. We have also examined the role of viscous diffusion in establishing steady state corotation torques in magnetically torqued discs that also support low level turbulent diffusion, and have developed expressions for these steady torques as a function of system parameters.

Finally, in Appendix~\ref{sec:prescription}, we have provided a prescription for including the dynamical corotation torques discussed in this paper in N-body simulations of planet formation and population synthesis models.

Future work suggested by these results includes the application of novel techniques for reducing numerical diffusion,
extension of the parameter regime to include less weakly coupled magnetic fields and transitions between disc parameter regimes, non-isothermal discs with stellar irradiation and internal radiative transfer,
three dimensional models, and N-body models of planetary system formation applying the torque formulae given in this paper. 

\section*{Acknowledgements}
This research was supported by STFC Consolidated grants awarded to the QMUL Astronomy Unit 2015-2018  ST/M001202/1 and 2017-2020 ST/P000592/1, and in part by the National Science Foundation under Grant No. NSF PHY17-48958.
This research utilised Queen Mary's Apocrita HPC facility, supported by QMUL Research-IT \citep{apocrita};
and the  DiRAC Data Centric system at Durham University, operated by the Institute for Computational Cosmology on behalf of the STFC DiRAC HPC Facility (www.dirac.ac.uk). 
This equipment was funded by a BIS National E-infrastructure capital grant ST/K00042X/1, STFC capital grant ST/K00087X/1, DiRAC Operations grant ST/K003267/1 and Durham University. DiRAC is part of the National E-Infrastructure.
SJP is supported by a Royal Society University Research Fellowship.




\bibliographystyle{mnras}

\begin{thebibliography}{}
\makeatletter
\relax
\def\mn@urlcharsother{\let\do\@makeother \do\$\do\&\do\#\do\^\do\_\do\%\do\~}
\def\mn@doi{\begingroup\mn@urlcharsother \@ifnextchar [ {\mn@doi@}
  {\mn@doi@[]}}
\def\mn@doi@[#1]#2{\def\@tempa{#1}\ifx\@tempa\@empty \href
  {http://dx.doi.org/#2} {doi:#2}\else \href {http://dx.doi.org/#2} {#1}\fi
  \endgroup}
\def\mn@eprint#1#2{\mn@eprint@#1:#2::\@nil}
\def\mn@eprint@arXiv#1{\href {http://arxiv.org/abs/#1} {{\tt arXiv:#1}}}
\def\mn@eprint@dblp#1{\href {http://dblp.uni-trier.de/rec/bibtex/#1.xml}
  {dblp:#1}}
\def\mn@eprint@#1:#2:#3:#4\@nil{\def\@tempa {#1}\def\@tempb {#2}\def\@tempc
  {#3}\ifx \@tempc \@empty \let \@tempc \@tempb \let \@tempb \@tempa \fi \ifx
  \@tempb \@empty \def\@tempb {arXiv}\fi \@ifundefined
  {mn@eprint@\@tempb}{\@tempb:\@tempc}{\expandafter \expandafter \csname
  mn@eprint@\@tempb\endcsname \expandafter{\@tempc}}}

\bibitem[\protect\citeauthoryear{{Bai}}{{Bai}}{2013}]{2013ApJ...772...96B}
{Bai} X.-N.,  2013, \mn@doi [\apj] {10.1088/0004-637X/772/2/96}, \href
  {http://adsabs.harvard.edu/abs/2013ApJ...772...96B} {772, 96}

\bibitem[\protect\citeauthoryear{{Bai}}{{Bai}}{2014a}]{2014ApJ...791...73B}
{Bai} X.-N.,  2014a, \mn@doi [\apj] {10.1088/0004-637X/791/1/73}, \href
  {http://adsabs.harvard.edu/abs/2014ApJ...791...73B} {791, 73}

\bibitem[\protect\citeauthoryear{{Bai}}{{Bai}}{2014b}]{2014ApJ...791..137B}
{Bai} X.-N.,  2014b, \mn@doi [\apj] {10.1088/0004-637X/791/2/137}, \href
  {http://adsabs.harvard.edu/abs/2014ApJ...791..137B} {791, 137}

\bibitem[\protect\citeauthoryear{{Bai}}{{Bai}}{2015}]{2015ApJ...798...84B}
{Bai} X.-N.,  2015, \mn@doi [\apj] {10.1088/0004-637X/798/2/84}, \href
  {http://adsabs.harvard.edu/abs/2015ApJ...798...84B} {798, 84}

\bibitem[\protect\citeauthoryear{{Bai}}{{Bai}}{2016}]{2016ApJ...821...80B}
{Bai} X.-N.,  2016, \mn@doi [\apj] {10.3847/0004-637X/821/2/80}, \href
  {http://adsabs.harvard.edu/abs/2016ApJ...821...80B} {821, 80}

\bibitem[\protect\citeauthoryear{{Bai} \& {Stone}}{{Bai} \&
  {Stone}}{2013}]{2013ApJ...769...76B}
{Bai} X.-N.,  {Stone} J.~M.,  2013, \mn@doi [\apj]
  {10.1088/0004-637X/769/1/76}, \href
  {http://adsabs.harvard.edu/abs/2013ApJ...769...76B} {769, 76}

\bibitem[\protect\citeauthoryear{{Bai}, {Ye}, {Goodman}  \& {Yuan}}{{Bai}
  et~al.}{2016}]{2016ApJ...818..152B}
{Bai} X.-N.,  {Ye} J.,  {Goodman} J.,   {Yuan} F.,  2016, \mn@doi [\apj]
  {10.3847/0004-637X/818/2/152}, \href
  {http://adsabs.harvard.edu/abs/2016ApJ...818..152B} {818, 152}

\bibitem[\protect\citeauthoryear{{Baruteau} \& {Masset}}{{Baruteau} \&
  {Masset}}{2008a}]{2008ApJ...672.1054B}
{Baruteau} C.,  {Masset} F.,  2008a, \mn@doi [\apj] {10.1086/523667}, \href
  {http://adsabs.harvard.edu/abs/2008ApJ...672.1054B} {672, 1054}

\bibitem[\protect\citeauthoryear{{Baruteau} \& {Masset}}{{Baruteau} \&
  {Masset}}{2008b}]{2008ApJ...678..483B}
{Baruteau} C.,  {Masset} F.,  2008b, \mn@doi [\apj] {10.1086/529487}, \href
  {http://adsabs.harvard.edu/abs/2008ApJ...678..483B} {678, 483}

\bibitem[\protect\citeauthoryear{{Ben{\'{\i}}tez Llambay} \&
  {Masset}}{{Ben{\'{\i}}tez Llambay} \& {Masset}}{2015}]{2015ascl.soft09006B}
{Ben{\'{\i}}tez Llambay} P.,  {Masset} F.,  2015, {FARGO3D:
  Hydrodynamics/magnetohydrodynamics code}, Astrophysics Source Code Library
  (\mn@eprint {ascl} {1509.006})

\bibitem[\protect\citeauthoryear{{Ben{\'{\i}}tez-Llambay} \&
  {Masset}}{{Ben{\'{\i}}tez-Llambay} \& {Masset}}{2016}]{2016ApJS..223...11B}
{Ben{\'{\i}}tez-Llambay} P.,  {Masset} F.~S.,  2016, \mn@doi [\apjs]
  {10.3847/0067-0049/223/1/11}, \href
  {http://adsabs.harvard.edu/abs/2016ApJS..223...11B} {223, 11}

\bibitem[\protect\citeauthoryear{{Ben{\'{\i}}tez-Llambay}, {Ramos},
  {Beaug{\'e}}  \& {Masset}}{{Ben{\'{\i}}tez-Llambay}
  et~al.}{2016}]{2016ApJ...826...13B}
{Ben{\'{\i}}tez-Llambay} P.,  {Ramos} X.~S.,  {Beaug{\'e}} C.,   {Masset}
  F.~S.,  2016, \mn@doi [\apj] {10.3847/0004-637X/826/1/13}, \href
  {http://adsabs.harvard.edu/abs/2016ApJ...826...13B} {826, 13}

\bibitem[\protect\citeauthoryear{{B{\'e}thune}, {Lesur}  \&
  {Ferreira}}{{B{\'e}thune} et~al.}{2017}]{2017A&A...600A..75B}
{B{\'e}thune} W.,  {Lesur} G.,   {Ferreira} J.,  2017, \mn@doi [\aap]
  {10.1051/0004-6361/201630056}, \href
  {http://adsabs.harvard.edu/abs/2017A%26A...600A..75B} {600, A75}

\bibitem[\protect\citeauthoryear{{Casoli} \& {Masset}}{{Casoli} \&
  {Masset}}{2009}]{2009ApJ...703..845C}
{Casoli} J.,  {Masset} F.~S.,  2009, \mn@doi [\apj]
  {10.1088/0004-637X/703/1/845}, \href
  {http://adsabs.harvard.edu/abs/2009ApJ...703..845C} {703, 845}

\bibitem[\protect\citeauthoryear{{Goldreich} \& {Tremaine}}{{Goldreich} \&
  {Tremaine}}{1979}]{1979ApJ...233..857G}
{Goldreich} P.,  {Tremaine} S.,  1979, \mn@doi [\apj] {10.1086/157448}, \href
  {http://esoads.eso.org/abs/1979ApJ...233..857G} {233, 857}

\bibitem[\protect\citeauthoryear{{Goodman} \& {Rafikov}}{{Goodman} \&
  {Rafikov}}{2001}]{Goodman2001}
{Goodman} J.,  {Rafikov} R.~R.,  2001, \mn@doi [\apj] {10.1086/320572}, \href
  {http://adsabs.harvard.edu/abs/2001ApJ...552..793G} {552, 793}

\bibitem[\protect\citeauthoryear{{Gressel}, {Turner}, {Nelson}  \&
  {McNally}}{{Gressel} et~al.}{2015}]{2015ApJ...801...84G}
{Gressel} O.,  {Turner} N.~J.,  {Nelson} R.~P.,   {McNally} C.~P.,  2015,
  \mn@doi [\apj] {10.1088/0004-637X/801/2/84}, \href
  {http://adsabs.harvard.edu/abs/2015ApJ...801...84G} {801, 84}

\bibitem[\protect\citeauthoryear{{Hourigan} \& {Ward}}{{Hourigan} \&
  {Ward}}{1984}]{Hourigan1984}
{Hourigan} K.,  {Ward} W.~R.,  1984, \mn@doi [\icarus]
  {10.1016/0019-1035(84)90136-2}, \href
  {http://adsabs.harvard.edu/abs/1984Icar...60...29H} {60, 29}

\bibitem[\protect\citeauthoryear{{King}, {Butcher}  \& {Zalewski}}{{King}
  et~al.}{2017}]{apocrita}
{King} T.,  {Butcher} S.,   {Zalewski} L.,  2017, Technical report, {Apocrita -
  High Performance Computing Cluster for Queen Mary University of London}.
Queen Mary University of London, \mn@doi{10.5281/zenodo.438045}

\bibitem[\protect\citeauthoryear{{Kunz}}{{Kunz}}{2008}]{2008MNRAS.385.1494K}
{Kunz} M.~W.,  2008, \mn@doi [\mnras] {10.1111/j.1365-2966.2008.12928.x}, \href
  {http://adsabs.harvard.edu/abs/2008MNRAS.385.1494K} {385, 1494}

\bibitem[\protect\citeauthoryear{{Kunz} \& {Lesur}}{{Kunz} \&
  {Lesur}}{2013}]{2013MNRAS.434.2295K}
{Kunz} M.~W.,  {Lesur} G.,  2013, \mn@doi [\mnras] {10.1093/mnras/stt1171},
  \href {http://adsabs.harvard.edu/abs/2013MNRAS.434.2295K} {434, 2295}

\bibitem[\protect\citeauthoryear{{Lambrechts}, {Johansen}  \&
  {Morbidelli}}{{Lambrechts} et~al.}{2014}]{Lambrechts2014}
{Lambrechts} M.,  {Johansen} A.,   {Morbidelli} A.,  2014, \mn@doi [\aap]
  {10.1051/0004-6361/201423814}, \href
  {http://adsabs.harvard.edu/abs/2014A%26A...572A..35L} {572, A35}

\bibitem[\protect\citeauthoryear{{Lesur}, {Kunz}  \& {Fromang}}{{Lesur}
  et~al.}{2014}]{2014A&A...566A..56L}
{Lesur} G.,  {Kunz} M.~W.,   {Fromang} S.,  2014, \mn@doi [\aap]
  {10.1051/0004-6361/201423660}, \href
  {http://adsabs.harvard.edu/abs/2014A%26A...566A..56L} {566, A56}

\bibitem[\protect\citeauthoryear{{Li}, {Lubow}, {Li}  \& {Lin}}{{Li}
  et~al.}{2009}]{Li2009}
{Li} H.,  {Lubow} S.~H.,  {Li} S.,   {Lin} D.~N.~C.,  2009, \mn@doi [\apjl]
  {10.1088/0004-637X/690/1/L52}, \href
  {http://adsabs.harvard.edu/abs/2009ApJ...690L..52L} {690, L52}

\bibitem[\protect\citeauthoryear{{Masset}}{{Masset}}{2001}]{2001ApJ...558..453M}
{Masset} F.~S.,  2001, \mn@doi [\apj] {10.1086/322446}, \href
  {http://adsabs.harvard.edu/abs/2001ApJ...558..453M} {558, 453}

\bibitem[\protect\citeauthoryear{{Masset} \& {Ben{\'{\i}}tez-Llambay}}{{Masset}
  \& {Ben{\'{\i}}tez-Llambay}}{2016}]{2016ApJ...817...19M}
{Masset} F.~S.,  {Ben{\'{\i}}tez-Llambay} P.,  2016, \mn@doi [\apj]
  {10.3847/0004-637X/817/1/19}, \href
  {http://adsabs.harvard.edu/abs/2016ApJ...817...19M} {817, 19}

\bibitem[\protect\citeauthoryear{{Masset} \& {Casoli}}{{Masset} \&
  {Casoli}}{2010}]{MassetCasoli2010}
{Masset} F.~S.,  {Casoli} J.,  2010, \mn@doi [\apj]
  {10.1088/0004-637X/723/2/1393}, \href
  {http://adsabs.harvard.edu/abs/2010ApJ...723.1393M} {723, 1393}

\bibitem[\protect\citeauthoryear{{Masset} \& {Papaloizou}}{{Masset} \&
  {Papaloizou}}{2003}]{2003ApJ...588..494M}
{Masset} F.~S.,  {Papaloizou} J.~C.~B.,  2003, \mn@doi [\apj] {10.1086/373892},
  \href {http://adsabs.harvard.edu/abs/2003ApJ...588..494M} {588, 494}

\bibitem[\protect\citeauthoryear{{Masset}, {D'Angelo}  \& {Kley}}{{Masset}
  et~al.}{2006}]{2006ApJ...652..730M}
{Masset} F.~S.,  {D'Angelo} G.,   {Kley} W.,  2006, \mn@doi [\apj]
  {10.1086/507515}, \href {http://adsabs.harvard.edu/abs/2006ApJ...652..730M}
  {652, 730}

\bibitem[\protect\citeauthoryear{{McNally}, {Nelson}, {Paardekooper}, {Gressel}
   \& {Lyra}}{{McNally} et~al.}{2017}]{2017MNRAS.472.1565M}
{McNally} C.~P.,  {Nelson} R.~P.,  {Paardekooper} S.-J.,  {Gressel} O.,
  {Lyra} W.,  2017, \mn@doi [\mnras] {10.1093/mnras/stx2136}, \href
  {http://adsabs.harvard.edu/abs/2017MNRAS.472.1565M} {472, 1565}

\bibitem[\protect\citeauthoryear{{Nelson}, {Gressel}  \& {Umurhan}}{{Nelson}
  et~al.}{2013}]{Nelson2013}
{Nelson} R.~P.,  {Gressel} O.,   {Umurhan} O.~M.,  2013, \mn@doi [\mnras]
  {10.1093/mnras/stt1475}, \href
  {http://adsabs.harvard.edu/abs/2013MNRAS.435.2610N} {435, 2610}

\bibitem[\protect\citeauthoryear{{Ogihara}, {Kokubo}, {Suzuki}, {Morbidelli}
  \& {Crida}}{{Ogihara} et~al.}{2017}]{2017arXiv171001240O}
{Ogihara} M.,  {Kokubo} E.,  {Suzuki} T.~K.,  {Morbidelli} A.,   {Crida} A.,
  2017, preprint, \href {http://adsabs.harvard.edu/abs/2017arXiv171001240O} {}
  (\mn@eprint {arXiv} {1710.01240})

\bibitem[\protect\citeauthoryear{{Paardekooper}}{{Paardekooper}}{2014}]{2014MNRAS.444.2031P}
{Paardekooper} S.-J.,  2014, \mn@doi [\mnras] {10.1093/mnras/stu1542}, \href
  {http://adsabs.harvard.edu/abs/2014MNRAS.444.2031P} {444, 2031}

\bibitem[\protect\citeauthoryear{{Paardekooper} \& {Mellema}}{{Paardekooper} \&
  {Mellema}}{2008}]{2008A&A...478..245P}
{Paardekooper} S.-J.,  {Mellema} G.,  2008, \mn@doi [\aap]
  {10.1051/0004-6361:20078592}, \href
  {http://adsabs.harvard.edu/abs/2008A%26A...478..245P} {478, 245}

\bibitem[\protect\citeauthoryear{{Paardekooper} \& {Papaloizou}}{{Paardekooper}
  \& {Papaloizou}}{2008}]{2008A&A...485..877P}
{Paardekooper} S.-J.,  {Papaloizou} J.~C.~B.,  2008, \mn@doi [\aap]
  {10.1051/0004-6361:20078702}, \href
  {http://adsabs.harvard.edu/abs/2008A%26A...485..877P} {485, 877}

\bibitem[\protect\citeauthoryear{{Paardekooper}, {Baruteau}, {Crida}  \&
  {Kley}}{{Paardekooper} et~al.}{2010}]{2010MNRAS.401.1950P}
{Paardekooper} S.-J.,  {Baruteau} C.,  {Crida} A.,   {Kley} W.,  2010, \mn@doi
  [\mnras] {10.1111/j.1365-2966.2009.15782.x}, \href
  {http://adsabs.harvard.edu/abs/2010MNRAS.401.1950P} {401, 1950}

\bibitem[\protect\citeauthoryear{{Paardekooper}, {Baruteau}  \&
  {Kley}}{{Paardekooper} et~al.}{2011}]{2011MNRAS.410..293P}
{Paardekooper} S.-J.,  {Baruteau} C.,   {Kley} W.,  2011, \mn@doi [\mnras]
  {10.1111/j.1365-2966.2010.17442.x}, \href
  {http://adsabs.harvard.edu/abs/2011MNRAS.410..293P} {410, 293}

\bibitem[\protect\citeauthoryear{{Pandey} \& {Wardle}}{{Pandey} \&
  {Wardle}}{2008}]{2008MNRAS.385.2269P}
{Pandey} B.~P.,  {Wardle} M.,  2008, \mn@doi [\mnras]
  {10.1111/j.1365-2966.2008.12998.x}, \href
  {http://adsabs.harvard.edu/abs/2008MNRAS.385.2269P} {385, 2269}

\bibitem[\protect\citeauthoryear{{Papaloizou}, {Nelson}, {Kley}, {Masset}  \&
  {Artymowicz}}{{Papaloizou} et~al.}{2007}]{2007prpl.conf..655P}
{Papaloizou} J.~C.~B.,  {Nelson} R.~P.,  {Kley} W.,  {Masset} F.~S.,
  {Artymowicz} P.,  2007, Protostars and Planets V, \href
  {http://adsabs.harvard.edu/abs/2007prpl.conf..655P} {pp 655--668}

\bibitem[\protect\citeauthoryear{{Pepli{\'n}ski}, {Artymowicz}  \&
  {Mellema}}{{Pepli{\'n}ski} et~al.}{2008a}]{2008MNRAS.386..164P}
{Pepli{\'n}ski} A.,  {Artymowicz} P.,   {Mellema} G.,  2008a, \mn@doi [\mnras]
  {10.1111/j.1365-2966.2008.13045.x}, \href
  {http://adsabs.harvard.edu/abs/2008MNRAS.386..164P} {386, 164}

\bibitem[\protect\citeauthoryear{{Pepli{\'n}ski}, {Artymowicz}  \&
  {Mellema}}{{Pepli{\'n}ski} et~al.}{2008b}]{2008MNRAS.386..179P}
{Pepli{\'n}ski} A.,  {Artymowicz} P.,   {Mellema} G.,  2008b, \mn@doi [\mnras]
  {10.1111/j.1365-2966.2008.13046.x}, \href
  {http://adsabs.harvard.edu/abs/2008MNRAS.386..179P} {386, 179}

\bibitem[\protect\citeauthoryear{{Pepli{\'n}ski}, {Artymowicz}  \&
  {Mellema}}{{Pepli{\'n}ski} et~al.}{2008c}]{2008MNRAS.387.1063P}
{Pepli{\'n}ski} A.,  {Artymowicz} P.,   {Mellema} G.,  2008c, \mn@doi [\mnras]
  {10.1111/j.1365-2966.2008.13339.x}, \href
  {http://adsabs.harvard.edu/abs/2008MNRAS.387.1063P} {387, 1063}

\bibitem[\protect\citeauthoryear{{Pierens}}{{Pierens}}{2015}]{2015MNRAS.454.2003P}
{Pierens} A.,  2015, \mn@doi [\mnras] {10.1093/mnras/stv2024}, \href
  {http://adsabs.harvard.edu/abs/2015MNRAS.454.2003P} {454, 2003}

\bibitem[\protect\citeauthoryear{{Pierens} \& {Raymond}}{{Pierens} \&
  {Raymond}}{2016}]{2016MNRAS.462.4130P}
{Pierens} A.,  {Raymond} S.~N.,  2016, \mn@doi [\mnras]
  {10.1093/mnras/stw1904}, \href
  {http://adsabs.harvard.edu/abs/2016MNRAS.462.4130P} {462, 4130}

\bibitem[\protect\citeauthoryear{{Rafikov}}{{Rafikov}}{2002}]{Rafikov2002}
{Rafikov} R.~R.,  2002, \mn@doi [\apj] {10.1086/340228}, \href
  {http://adsabs.harvard.edu/abs/2002ApJ...572..566R} {572, 566}

\bibitem[\protect\citeauthoryear{{Simon}, {Lesur}, {Kunz}  \&
  {Armitage}}{{Simon} et~al.}{2015a}]{2015MNRAS.454.1117S}
{Simon} J.~B.,  {Lesur} G.,  {Kunz} M.~W.,   {Armitage} P.~J.,  2015a, \mn@doi
  [\mnras] {10.1093/mnras/stv2070}, \href
  {http://adsabs.harvard.edu/abs/2015MNRAS.454.1117S} {454, 1117}

\bibitem[\protect\citeauthoryear{{Simon}, {Lesur}, {Kunz}  \&
  {Armitage}}{{Simon} et~al.}{2015b}]{Simon2015}
{Simon} J.~B.,  {Lesur} G.,  {Kunz} M.~W.,   {Armitage} P.~J.,  2015b, \mn@doi
  [\mnras] {10.1093/mnras/stv2070}, \href
  {http://adsabs.harvard.edu/abs/2015MNRAS.454.1117S} {454, 1117}

\bibitem[\protect\citeauthoryear{{Ward}}{{Ward}}{1991}]{1991LPI....22.1463W}
{Ward} W.~R.,  1991, in Lunar and Planetary Science Conference.

\bibitem[\protect\citeauthoryear{{Wardle} \& {Ng}}{{Wardle} \&
  {Ng}}{1999}]{1999MNRAS.303..239W}
{Wardle} M.,  {Ng} C.,  1999, \mn@doi [\mnras]
  {10.1046/j.1365-8711.1999.02211.x}, \href
  {http://adsabs.harvard.edu/abs/1999MNRAS.303..239W} {303, 239}

\bibitem[\protect\citeauthoryear{{de Val-Borro} et~al.,}{{de Val-Borro}
  et~al.}{2006}]{2006MNRAS.370..529D}
{de Val-Borro} M.,  et~al., 2006, \mn@doi [\mnras]
  {10.1111/j.1365-2966.2006.10488.x}, \href
  {http://adsabs.harvard.edu/abs/2006MNRAS.370..529D} {370, 529}

\makeatother
\end{thebibliography}



\appendix
\section{Radially refined grid}
\label{sec:refinedgrid}

\begin{figure}
\includegraphics[width=\columnwidth]{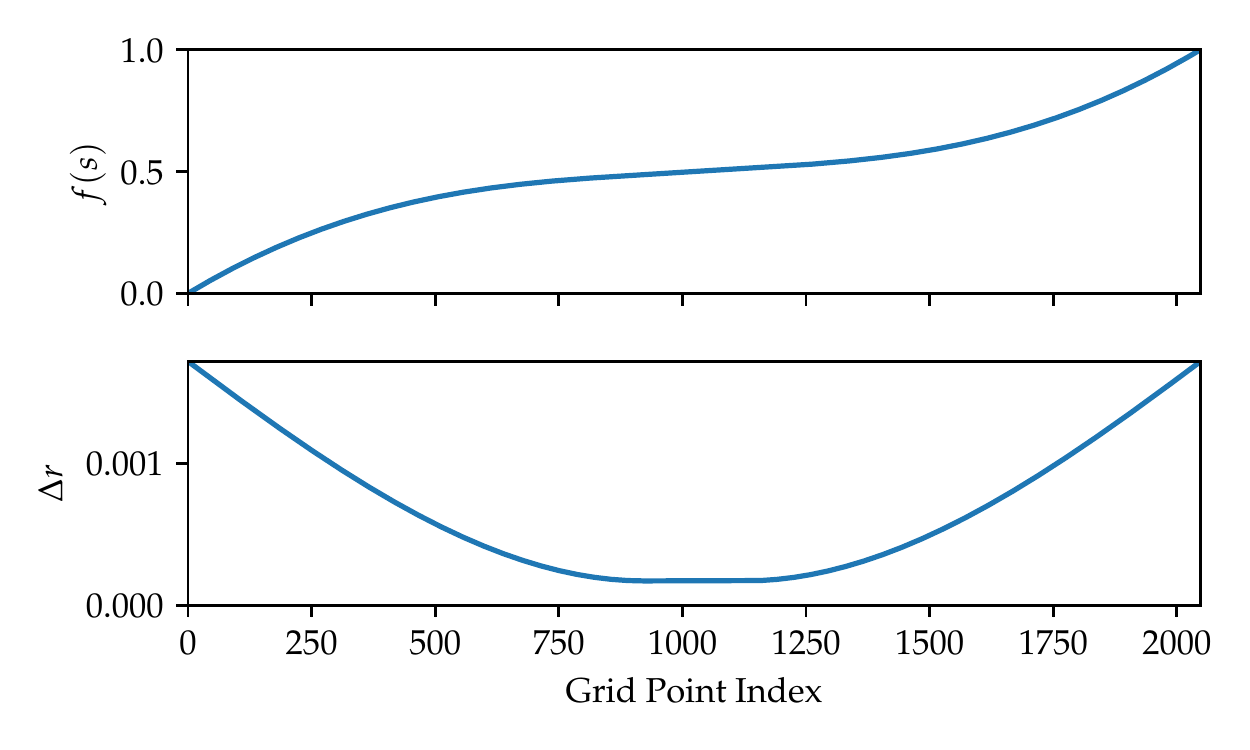}
\caption{Radially refined grid. Top: Mapping function $f(s)$, Bottom: Grid point radial spacing $\Delta r$.}
\label{fig:refinedgrid}
\end{figure}

%
%
%

We choose a core region of with $c=h=0.05$ to each side of the planet,
and grid it at a constant resolution $\Delta r$,
we then gradually increase the grid spacing outside that region, following a forth-order polynomial for the grid positions, thus 
matching the grid spacing at the join to the inner high resolution region and having zero first, second, and third derivatives of the grid spacing at the join.

Then, for a list of $1024$ evenly spaced points in the interval  $[0,1/2]$ the mapping which produces the grid is

\begin{align}
 f(s) = \begin{cases}
 a_4 x^4 + a_3 x^3 + a_2 x^2 + a_1 x + a_0 \quad &\text{if\ } s<1/2-c  \\
 S_c(s-1/2)+1/2\quad& \text{otherwise} \\
 \end{cases}
\end{align}
which specifies a linearly spaced core grid of half-width $c$ transitioning into a variably spaced grid 
where the generating polynomial coefficients are
\begin{align}
a_0 &= 0\\
a_1 &= M S_c\\
a_2 &= (1-2 c)^{-2} \left[ 6 (2+(2 c(M-1) -M-1)S_c)\right]\\
a_3 &= (2c -1)^{-3} \left[ 4 (8 +(6 c(M-1)-3M-5)S_c) \right]\\
a_4 &= (1-2c)^{-4}  \left[ 8 (3 +(2c(M-1) -M-2 ) S_c)\right]
\end{align}
which is a polynomial with slope $S_c$ where it joins the core grid, and slope $M S_c$ at the edge ($s=0$).
We use $S_c=1/4$ and $M=10$. The final grid point positions are the mapping of this half-interval to half of the domain with
\begin{align}
r(f(s)) = r_1+(r_1+r_2)f(s)
\end{align}
where $r_1= 1.70273437$ and $r_2= 0.29726562$, so that the above specifies the grid points at $r<1$. 
The outer half of the grid where $r>1$ is specified symmetrically about $r=1$, as shown in Figure~\ref{fig:refinedgrid}.

\section{Migration torque prescription for laminar magnetically torqued discs}
\label{sec:prescription}
Here we outline a simple approach to incorporating the corotation torque in a laminar, magnetically torqued disc in N-body simulations of planet formation or population synthesis calculations. We assume that the Lindblad torque will be included using a torque formula such as may be obtained from \cite{2010MNRAS.401.1950P}.
The two most likely regimes for a planet to be in are the regimes (i) and (ii) discussed in Section~\ref{sec:basicanal}, which both apply to a disc that sustains an inwards accretion flow onto the star, and here we focus on modelling the corotation torque in these two regimes. 

We assume that the disc model has a power-law surface density profile of the form
\begin{equation}
\Sigma(r) = \Sigma_0 \left(\frac{r}{r_0}\right)^{-\alpha},
\label{eq:Sigma_r}
\end{equation}
and that the time dependent mass accretion rate is constant at each radius and is given by
\begin{equation}
{\dot m}_{\rm HS}(t) = -2 \pi r \Sigma(r) v_r(r),
\label{eq:mdot}
\end{equation}
where $v_r(r)$ is the radius dependent radial velocity associated with the magnetically driven accretion flow. The accretion rate given in equation~(\ref{eq:mdot}) is the component driven by the Hall-effect-induced horizontal magnetic fields, and does not include the contribution due to the launching of a magnetised wind near the disc surface. This surface component does not contribute to the evolution of the corotation torque in our model, but it would be relatively trivial to include its effects on the evolution of the disc's mass budget if so desired.
For such other unspecified accretion sources we include a second mass accretion term ${\dot m}_{\rm x}$ in the following.

If the disc has fixed values of the inner and outer radii, $r_{\rm in}$ and $r_{\rm out}$, such that we ignore any evolution of these due to the torques that may be applied to them, then at each moment in time we can write
\begin{equation}
\Sigma_0(t) = \frac{(2-\alpha) \, m_{\rm d} (t)}{2 \pi r_0^{\alpha} \left(r_{\rm out}^{2-\alpha} - r_{\rm in}^{2-\alpha}\right)}.
\label{eq:Sigma_0}
\end{equation}
We assume that a model is computed by stepping forward in time using discrete time steps, $\Delta t$, such that $t_{i+1} =  t_i+\Delta t$. In the absence of any matter sources, the disc accretes onto the star and the total disc mass evolves according to 
\begin{equation}
m_{\rm d}(t_{i+1}) = m_{\rm d}(t_i) - ({\dot m}_{\rm HS} + {\dot m}_{\rm x} )\Delta t.
\label{eq:m_d}
\end{equation}

In principle, one has a choice about how to treat the global, time dependent mass accretion rate through the disc. One choice could be to specify how ${\dot m_{\rm HS}}(t)$ and  ${\dot m}_{\rm x}(t)$ vary with time and to obtain $v_r(r)$ by rearranging equation~(\ref{eq:mdot}). Another choice could be to assume that $v_r(r)$ is time independent, ${\dot m}_{\rm x}(t)=0$,  and allow the mass accretion rate to be determined by equation~(\ref{eq:mdot}). Here, for simplicity, we adopt the second of these choices such that ${\dot m}_{\rm HS}$ decreases with time as $m_{\rm d}$ and $\Sigma$ decrease, since we assume that no mass is added to the disc during its life time.

For a Keplerian disc we have the following relation between the radial velocity associated with the accretion flow and the azimuthal acceleration, $a_{\phi}$, applied to the disc due to the Lorentz force
\begin{equation}
v_r = \frac{2 r^2 a_{\phi}}{\sqrt{G M_* r}}.
\label{eq:v_r}
\end{equation}
We note that the Lorentz acceleration acting on a two dimensional disc can be written
\begin{equation}
a_{\phi} = \frac{\langle {\bf J} \times {\bf B} \rangle_{\phi}}{\Sigma},
\label{eq:a_phiinb}
\end{equation}
where ${\bf J}$ is the current density, which may be expressed as ${\bf J}=(\nabla \times {\bf B})/4 \pi$, and ${\bf B}$ is the magnetic field. The angled brackets indicate that the Lorentz force has been vertically averaged. When performing simple disc modelling, however, knowledge of the magnetic field strength and current density are not required. All that is needed to define how the disc and corotation torque evolves is knowledge of the torque that is applied to the disc, and not its mathematical form or physical origin\footnote{If one wants to include the dissipation of magnetic fields via Joule heating in the energy balance of the disc, then clearly a more sophisticated model that includes knowledge of the magnetic fields and currents would be required.}. By defining an initial mass accretion rate and an initial surface density profile, the required value of $v_r$ can be determined from equation~(\ref{eq:mdot}), and this then defines the value of $a_{\phi}$ that must operate in the disc. From equation~(\ref{eq:v_r}) we have
\begin{equation}
a_{\phi} = \frac{\sqrt{G M_* r}}{2 r^2} v_r.
\label{eq:a_phi}
\end{equation}
This is used below to calculate the evolution of the corotation torque.

\subsection{Corotation torque in regime (i)}
\label{sec:Regime_i}
We recall that regime~(i) applies when both planet and disc gas are drifting inwards, and the planet is drifting faster than the gas. Quoting from equation~(\ref{eq:unified}) in Section~\ref{sec:basicanal}, the corotation torque in regime~(i) should be computed according to
\begin{equation}
\Gamma_{\rm hs} = 2\pi \left( 1-\frac{\wc (t)}{w(\rp)}\right) \Sigma_{\rm p} \rp ^2 x_s \Omega_{\rm p} \left[\frac{d \rp}{dt} -v_r \right],
\label{eq:Gamma_hs}
\end{equation}
where $\wc$ and $w(\rp)$ are the inverse vortensities in the corotation region and in the background disc at the planet's location, respectively, and subscript `p' indicates that a quantity should be evaluated at the planet's location. As demonstrated in \citetalias{2017MNRAS.472.1565M} (see equation~(29) in Section~2.2.2 of that paper), the accretion-driving magnetic torque causes $\wc(t)$ to evolve according to
\begin{equation}
\frac{d\wc}{dt}= - \wc^2 \left[ \frac{1}{\Sigma r} \frac{d}{dr}\left(r a_{\phi}\right) \right].
\label{eq:dwdt}
\end{equation}
For a model that is evolving via discrete time steps, the solution to equation~(\ref{eq:dwdt}) that can be used to evolve $\wc(t)$ can be written
\begin{equation}
\wc(t_{i+1}) = \wc(t_i) \left[ 1 + \left(\frac{3}{2} - \alpha \right) \frac{{\dot m}_{\rm HS} \, \Omega_{\rm p} \wc(t_i)}{4 \pi \Sigma_{\rm p}^2 r_{\rm p}^2} \Delta t \right]^{-1}.
\label{eq:wc}
\end{equation}
Integration of equation~(\ref{eq:wc}) can be initiated at time $t=0$ by setting $\wc(0)=w(\rp)$, such that the inverse vortensity in the corotation region initially has the same value as the surrounding disc. The motion of the planet through the disc causes $w(\rp)$ to evolve since for a Keplerian disc we have
\begin{equation}
w(\rp) = \frac{2\Sigma_{\rm p}}{\Omega_{\rm p}}.
\label{eq:w_rp}
\end{equation}
In summary, when in regime (i) one can use equation~(\ref{eq:Gamma_hs}) to calculate the corotation torque as a function of time, using equation~(\ref{eq:wc}) to update $\wc$, and equation~(\ref{eq:w_rp}) to update $w(r_{\rm p})$ as the planet migrates and the disc is torqued. At $t=0$, the corotation torque $\Gamma_{\rm hs}=0$ because $\wc=w(\rp)$, and then $dr_{\rm p}/dt$ is determined by the Lindblad torque. In regime (i), when $|dr_{\rm p}/dt| > |v_r|$, the migration will be slowed as the corotation torque evolves until $d r_{\rm p}/dt \sim v_r$.

\subsection{Corotation torque in regime (ii)}
We recall that regime (ii) corresponds to both planet and disc drifting inwards initially, but with the disc drifting in faster than the planet. Here, we expect the planet to slow down, stop, reverse its migration, undergo a period of accelerating outward migration, and to then settle into fast outward migration at a steady speed. As discussed in Section~\ref{sec:fastmigration}, we do not have an analytical prediction for what the torque should be during the fast steady phase of outward migration, since saturation of the runaway is caused by complex, nonlinear gas flows. Hence, to obtain an expression for this torque we adopt a simple formula that interpolates between the steady torque values obtained in the suite of runs described in Section~\ref{sec:fastnumerical} and presented in figure~\ref{fig:scanfast}. The torque, as a function of the surface density power law in the disc and the flow-through parameter, $\chi_{\rm G}$, is given by
\begin{align}
\label{eq:Gamma_fast}
\frac{\Gamma_{\rm hs}(\chi_{\rm G}, \alpha) }{\Gamma_0}=& \frac{b_{j+1}-\alpha}{b_{j+1}-b_j} \left( \frac{a_{i+1} - \chi_{\rm G}}{a_{i+1} - a_i} Q_{i, j} + \frac{ \chi_{\rm G} -a_i}{a_{i+1} - a_i} Q_{i+1, j} \right) \nonumber \\
+ \frac{\alpha - b_{j}}{b_{j+1}-b_j}&\left(   \frac{a_{i+1} - \chi_{\rm G}}{a_{i+1} - a_i} Q_{i, j+1} + \frac{ \chi_{\rm G} -a_i}{a_{i+1} - a_i} Q_{i+1, j+1}  \right) , \\
i=&\begin{cases} 
 1 \text{ if } 0.0 \leq \chi_{\rm G} < 0.2  \\
 2 \text{ if } 0.2 \leq \chi_{\rm G} < 0.8 \\
 3 \text{ if } 0.8 \leq \chi_{\rm G} < 1.0 
\end{cases},\nonumber \ 
j=\begin{cases}
 1 \text{ if } 1.5 \geq \alpha > 1.0  \\
 2 \text{ if } 1.0 \geq \alpha > 0.5 \\
 3 \text{ if } 0.5 \geq \alpha > 0.0 
\end{cases},\nonumber\\
\mathbf{a} =& 
\begin{bmatrix}
0.0 \\
0.2\\
0.8\\
1.0
\end{bmatrix}\nonumber , \quad
\mathbf{b} =
\begin{bmatrix}
1.5 \\
1.0\\
0.5\\
0.0
\end{bmatrix}\nonumber , \quad
\mathbf{Q} = 
\begin{bmatrix}
0.0 & 0.0 & 0.0 &0.0 \\
0.0 & 2.3 & 3.1 & 3.7 \\
0.0 & 5.8 & 6.7 & 7.3 \\
0.0 & 6.1 & 6.9 & 7.5 
\end{bmatrix}\nonumber ,
\end{align}
where, in accordance with conventional mathematical notation the indices of the matrix $\mathbf{Q}$ are in the order  row, column 
(which differs from the array index notation of some programming languages).
We recall that
\begin{equation}
\chi_{\rm G} = \frac{3 x_s^2 \Omega_{\rm p}}{4 \pi r_{\rm p} (dr_{\rm p}/dt - v_r)}\ .
\end{equation}

We would expect that in a simulation of planet migration in a torqued inviscid disc that uses torque formulae such as those presented here to compute a planet's orbital evolution, at the beginning of the calculation $\Gamma_{\rm hs}=0$ and the planet will migrate inwards at a rate determined by the Lindblad torque. If the planet is in regime (ii), then the corotation torque determined by equation~(\ref{eq:Gamma_hs}) will increase, and will cause the planet to reverse its migration and enter a runaway phase. The runaway will saturate, however, when the corotation torques predicted by equations~(\ref{eq:Gamma_hs}) and~(\ref{eq:Gamma_fast}) are equal, and at this point one should calculate the steady torque given by equation~(\ref{eq:Gamma_fast}).

\subsection{Evolution with changing planet mass}
\label{sec:planetmass}
So far we have only discussed how to calculate torques for planets of fixed mass, but to be useful in planet formation simulations we need to adjust the corotation torque when the planet gains mass because this widens the horseshoe region, which allows gas from the background disc enter the corotation region and start librating. The vortensity of this newly librating gas will clearly differ from that which was originally in the horseshoe region, and this will affect the corotation torque. To account for this, it is reasonable to assume that there will be mixing of material in the horseshoe region, and the value of $\wc$ should then be modified each time the planet mass is incremented such that it becomes an area weighted average of the original value of $\wc$ and $w(\rp)$.

 \section{Model for Regimes (\texorpdfstring{\MakeLowercase{i}}{i}) and (\texorpdfstring{\MakeLowercase{iii}}{iii}) with viscosity}
\label{sec:viscmodel}
The evolution of the corotation region in a laminar magnetically torqued disc leads to a sharp top-hat distribution of vortensity in the libration island, as shown in \citetalias{2017MNRAS.472.1565M}. Here we extend the approach taken in \citet{2014MNRAS.444.2031P} that accounts for a finite turbulent viscosity in the disc, and examine the steady state corotation torques that can that arise in the presence of vortensity diffusion. We first develop an analytical approximation, and then verify its application with a set of simulations.
A similar final torque formula has been conjectured by \citet{2017arXiv171001240O}, although credited to \citet{2014MNRAS.444.2031P}. In this section we make it clear how the model which generates this torque formula 
differs from the one given in \citet{2014MNRAS.444.2031P}.

\subsection{Analytical theory with viscosity}
Before presenting the derivation, we note that this section develops a model for regime~(i) and (iii) with a finite viscosity only, as the canonical
 model for the inviscid case in these regimes at late times is that the radial migration of the planet is simply the same as the radial motion of gas in the disc.

Our model for the case of finite viscosity follows from the ones of \citet{2011MNRAS.410..293P} and \citet{2014MNRAS.444.2031P}.
The evolution of the inverse vortensity, $w$, in the corotation region is modelled with a one-dimensional viscous disc equation.
Here, the radial scaling of the viscosity in assumed to be such that it does not drive accretion ($\nu\propto r^{\alpha-1/2}$) but as the model only applies locally
at the corotation region on the length scale $x_s$ (the half-width of the corotation region) the radial gradient does not significantly affect the results,  so that it can be used for general viscosity power laws.

In terms of the nondimensional radial coordinate $\bar{x} = (r-\rp)/\rp$, the evolution of the inverse vortensity $w$ 
in the disc is
\begin{align}
\frac{\partial w(\bar{x},t) }{\partial t} =& \frac{3 \nup}{ \rp^2}(1+\bar{x})^{1/2} \frac{\partial }{\partial \bar{x}}
  \left( (1+\bar{x})^{1/2} \frac{\partial }{\partial \bar{x}}\left( (1+\bar{x})^{\alpha -3/2} w(\bar{x},t) \right)\right) \nonumber\\
 & -\frac{d \rp}{dt} \left(\frac{3}{2} - \alpha \right) \frac{\wwp}{\rp} (1+\bar{x})^{1/2-\alpha} \Pi\left(\frac{\bar{x}}{\bar{x}_s}\right)\nonumber\\
 & - w^2 \left[\frac{1}{\Sigma}  \nabla \times \frac{ T}{r\Sigma} \right] \Pi\left(\frac{\bar{x}}{\bar{x}_s}\right) \ ,
 \label{eq:wdiffeq}
\end{align}
where $\Pi$ is the rectangular function ( $1 \in [-1,1]$ and $0$ otherwise), $T$ is the accretion torque exerted on the gas by Maxwell stresses, 
and $\nup$ is the kinematic viscosity of the disc at the planet radius.
The body force, or Lorentz force of the laminar magnetic field, is included as the final term in equation~(\ref{eq:wdiffeq}).
Equation~(\ref{eq:wdiffeq}) can thus also be read as an extension to the model for the inviscid corotation region vortensity given in equation~(29) of \citetalias{2017MNRAS.472.1565M}.
Note that this model neglects all radial advection of vortensity, and thus 
consistently neglects the magnetic torque body force outside of the corotation region.
As in \citet{2014MNRAS.444.2031P} we can seek a steady-state solution for $w(\bar{x})$
in the corotation region.

The final driving term of equation~(\ref{eq:wdiffeq}) can be rewritten in terms of the steady-state inflow velocity driven by the magnetic field $v_r$
\begin{align}
 - w^2  \left(\frac{3}{2} -\alpha \right) \frac{(-v_r)  }{\rp \wwp} \left( 1 + \bar{x} \right)^{-5/2+\alpha},
\end{align}
where again $\bar{x} = (r-\rp)/\rp$.
Using this form in equation~(\ref{eq:wdiffeq}) we then change to the variables $z= 2(1+\bar{x})^{1/2} -2$ and 
$f = (1+z/2)^{2\alpha-3} w(\bar{x},t)$, and seek steady-state solutions for $f(z,t)$, which yields:
\begin{align}
0
=& \frac{d^2 f }{dz^2} \nonumber\\
&- \left[ \frac{d \rp}{dt}\ffp + (-v_r)\frac{f^2}{\ffp} \right]\left( \frac{3}{2} - \alpha\right) \frac{\rp }{3 \nup} \left( 1+z/2 \right)^{1-2\alpha}   \Pi\left(\frac{z}{z_s}\right) \ ,
\end{align}
where $z_s$ is $z(x_s)$.
It is now apparent where Galilean invariance of the torque will arise from in the expression in square brackets.
Now, expanding to leading (zeroth) order in $z\ll1$, as the corotation region width is small compared to the semimajor axis of the planet's orbit, gives:
\begin{align}
0 =& \frac{d^2 f }{dz^2} - \left[ \frac{d \rp}{dt}\ffp + (-v_r)\frac{f^2}{\ffp} \right]\left( \frac{3}{2} - \alpha\right) \frac{\rp }{3 \nup}  \Pi\left(\frac{z}{z_s}\right) \, .
\end{align}
We can then write the solution as $f(z,t) = \ffp+f_1(z,t)$ where $f_{\rm p}$ is the unperturbed value of $f$ at the planet's position 
and expand in $f_1\ll f_p$ (as the perturbation in vortensity due to the planet is small) to leading order which gives
\begin{align}
0 =& \frac{d^2 f_1}{dz^2} - \left[ \frac{d \rp}{dt} -v_r \right] \left( \frac{3}{2} - \alpha\right) \frac{\rp \ffp }{3 \nup}  \Pi\left(\frac{z}{z_s}\right) \ .
\end{align}
This is essentially the same equation as \citet{2014MNRAS.444.2031P} is solving to get his equation~(27) (except that our boundary condition is $f_1(-z_s)=f_1(z_s)=0$), 
so we only give here the new solution in the corotation region, which is
\begin{align}
\frac{f(z)}{\ffp} = 1+\left(\frac{3}{2}-\alpha\right) \frac{\rp}{3 \nup} \left[ \frac{d \rp}{dt} -v_r \right]\frac{z^2-z_s^2}{2}\ .
\end{align}
Taking $f(0)=\wc$ leads to 
\begin{align}
1-\frac{\wc}{w(\rp)} = \left(\frac{3}{2}-\alpha\right) \frac{x_s^2}{6 \rp \nup} \left[ \frac{d \rp}{dt} -v_r \right]  \ .
\label{eq:wcapproxvisc}
\end{align}
Substituting this model of the inverse vortensity of the corotation region into equation~(\ref{eq:unified}) gives a steady-state horseshoe torque for regime~(i) and regime~(iii) migration as
\begin{align}
\Gamma_{{\rm hs},\nu} = 2\mpi \left(\frac{3}{2}-\alpha\right) \frac{x_s^3}{6 \nup} \Sigma_{\rm p} \rp \Omega_{\rm p} \left[ \frac{d \rp}{dt} -v_r\right]^2\ .
\label{eq:gammahsnu}
\end{align}
This form again encodes the Galilean invariance property of the dynamic corotation torque, in that the final velocity term is the 
relative disc gas-planet radial velocity.
We now proceed to test this model with numerical simulations.

\subsection{Numerical tests of viscous regime (i) and (iii)}

\begin{figure}
\includegraphics[width=\columnwidth]{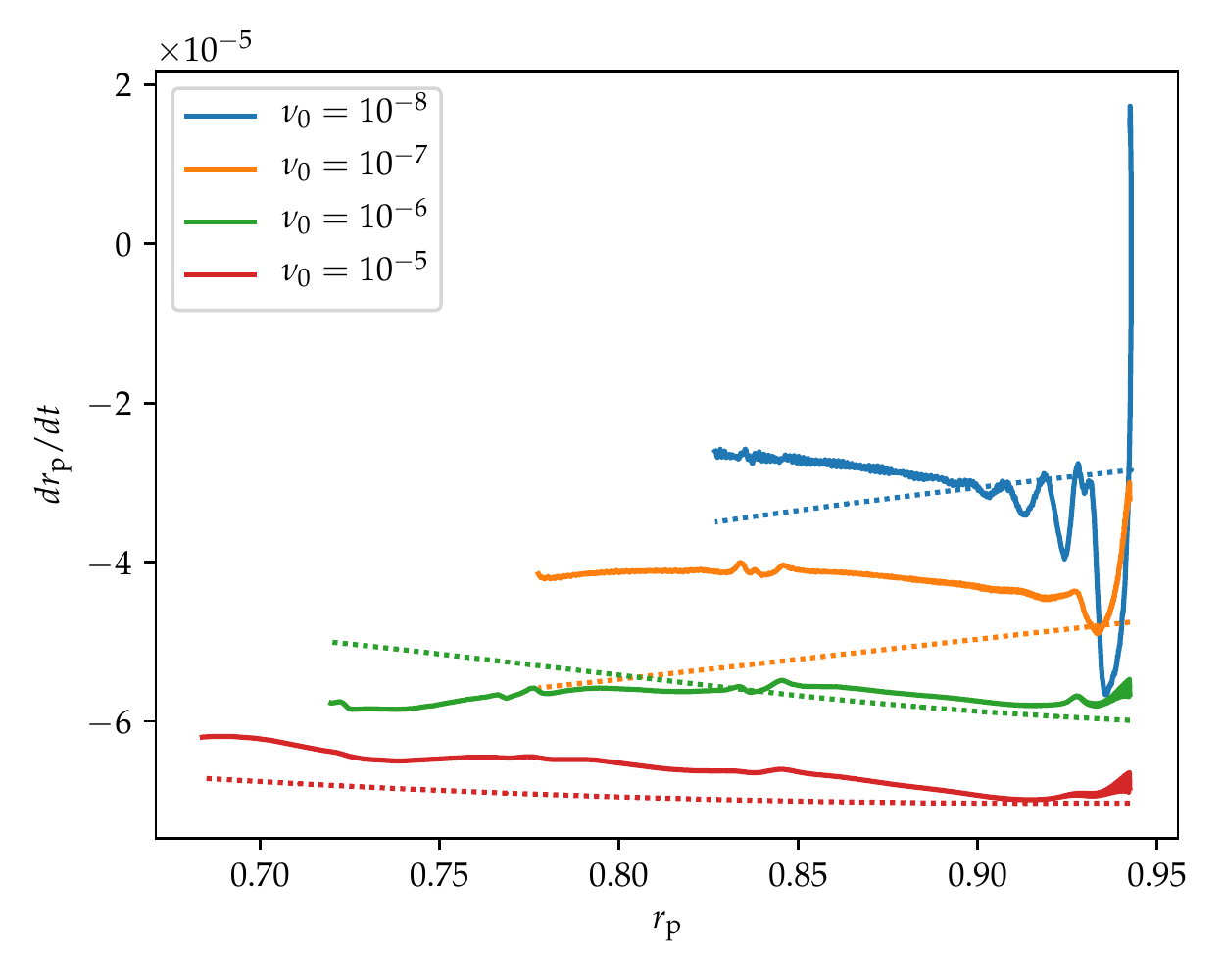}
\caption{Regime (i) with finite viscosity
$2\tau_{\rm lib}$ to $10\tau_{\rm lib}$
Theoretical steady-state migration rate given by the corresponding dotted line
}
\label{fig:regimeivisc}
\end{figure}

\begin{figure}
\includegraphics[width=\columnwidth]{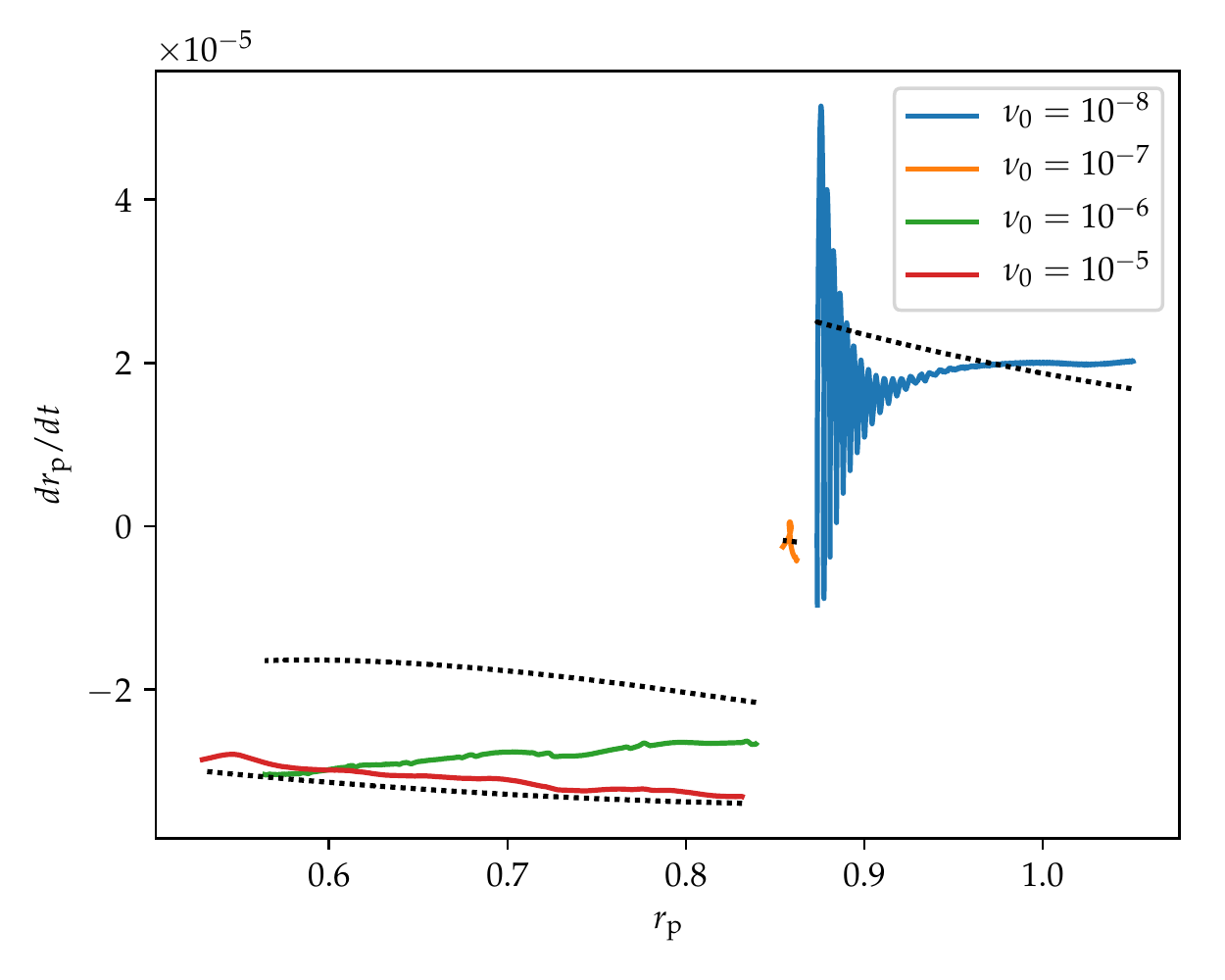}
\caption{Regime (iii) with finite viscosity shown from $4\tau_{\rm lib}$ to $24\tau_{\rm lib}$ 
Theoretical steady-state migration rates given by the black dotted lines
}
\label{fig:regimeiiivisc}
\end{figure}

\begin{table}
\caption{Characteristic values of parameters $\kappa$, $p$ and $\chi$ in tests}
\label{tab:sstimes}
\begin{center}
\begin{tabular}{clllr}
Migration regime & $\nu_0$ & $\kappa$ &$p$ & $\chi_{\rm G}$ \\
\hline
Regime (i)                            & $10^{-8}$ & $10^{-2}$ & $6$ & $5.6$ \\
Figure~\ref{fig:regimeivisc}  & $10^{-7}$ & $10^{-1}$ & $2$ & $3.1$\\
                                            & $10^{-6}$ & $10^{0}$   & $0.6$&  $2.1$\\
                                            & $10^{-5}$ & $10^{1}$   & $0.2$& $1.6$\\
\\
Regime (iii)                          & $10^{-8}$ & $10^{-2}$  & $6$  & $-5.9$\\  
Figure~\ref{fig:regimeiiivisc}& $10^{-7}$ & $10^{-1}$  & $2$  & $-2.6$ \\
                                            & $10^{-6}$ & $10^{1}$   & $0.6$ & $-1.6$\\
                                            & $10^{-5}$ & $10^{2}$   & $0.2$& $-1.4$\\
\end{tabular}
\end{center}
\end{table}

To demonstrate the level of agreement between the steady-state analytical model equation~(\ref{eq:gammahsnu}) and numerical simulations 
we present here a series of tests run in the same the fiducial domain as in Section~\ref{sec:specifiedmigration} 
and resolution $(N_r,N_\phi) = (1024,2048)$.
Again as in Section~\ref{sec:viscous} we use the surface density scaling $\Sigma = \Sigma_0 (r/r_0)^{-1/2}$ 
and viscosity radial scaling $\nu=\nu_0 (r/r_0)^{1/2}$.
First we ran a series of tests with a 
 magnetic inflow torque producing
$\chi=5$,
surface density 
$\Sigma_0 =6.3362\times10^{-3}$,
and viscosities
$\nu_0=10^{-5}$, $10^{-6}$, $10^{-7}$ and $10^{-8}$.
The resulting migration rates are shown in 
Figure~\ref{fig:regimeivisc}
along with the steady-state migration rate predicted by the total toque $\Gamma$ modelled as:
\begin{align}
\Gamma = \Gamma_{\rm L} + \Gamma_{\rm c,baro} + \Gamma_{{\rm hs},\nu} \ .  \label{eq:gammatotsteady}
\end{align}
where $\Gamma_{\rm L}$ is the Lindblad torque given by \citet{2011MNRAS.410..293P} their equation~(3),
 $\Gamma_{\rm c,baro}$ is the barotropic corotation torque given by \citet{2011MNRAS.410..293P} their equation~(32),
 and $\Gamma_{{\rm hs},\nu}$ is given by equation~(\ref{eq:gammahsnu}).
At the largest viscosity, the dynamical corotation torque contribution is very small, and the corotation torque is largely unsaturated,
so that the total torque is essentially the linearly predicted Lindblad and unsaturated corotation torques.
As the viscosity is decreased the relative contribution of the dynamical corotation torque increases.
At the lower viscosities one can also see the impact of libration oscillations on the planet motion; an initial transient which 
dies away as the libration island becomes well mixed.

A second scan of viscosities was run with a midplane outflow torque,
producing $\chi=-2$ with disc surface density $\Sigma_0 =3.1831\times10^{-3}$.
The resulting migration rates and predictions of equation~(\ref{eq:gammatotsteady}) for viscosities
$\nu_0=10^{-5}$, $10^{-6}$, $10^{-7},$ and $10^{-8}$
are shown in 
Figure~\ref{fig:regimeiiivisc}.
Again at the lowest $\nu_0=10^{-8}$ viscosity the libration oscillations are very apparent,
but equation~(\ref{eq:gammatotsteady}) does a quite reasonable job at predicting the migration rate
and the transition from an unsaturated linear corotation torque to a dynamical corotation torque.

The theory leading to equation~(\ref{eq:gammahsnu}) has formally limited applicability when the 
corotation region vortensity evolution is not dominantly a diffusion problem with time-constant source terms.
To quantify this one can evaluate
 the ratio between the crossing time of the radial flow across the corotation 
region (in the planet's frame) and the diffusion timescale across the corotation region.
For convenience we define this ratio as $\kappa$
\begin{align}
\kappa = \frac{x_s/(|v_r+d\rp/dt|)}{x_s^2/\nu}
\end{align}
and tabulate the characteristic values for each set of parameters in 
Figures~\ref{fig:regimeivisc} and~\ref{fig:regimeiiivisc} in Table~\ref{tab:sstimes}. 
For the lower values of viscosity, $\kappa$ is less than unity, indicating the the torque model would not be expected to preform well.
However, the torque model generally follows the correct trend,
suggesting that it does a reasonable job of including the relevant effects.
We also list in Table~\ref{tab:sstimes} the values of the parameter controlling viscous unsaturation of the corotation torque
 $p$ as defined in \citet{2011MNRAS.410..293P}, where for  $p \lesssim 1$ the corotation torque is unsaturated by viscosity.
When this is the case, the dynamical corotation torque effect is also negligible, as the vortensity in the corotation region 
simply acquires a gradient matching the background disk flow.
In this case the migration rate from equation~(\ref{eq:gammatotsteady}) is essentially as given by the  \citet{2011MNRAS.410..293P} formulae, 
with increasingly little contribution from dynamical torques.


\bsp	
\label{lastpage}
\end{document}